\documentclass[a4paper,11pt]{article}
\pdfoutput=1
\usepackage[english]{babel}
\usepackage{jheppub}
\pdfoutput=1
\usepackage{amsmath,amssymb,graphicx,xspace,slashed,epstopdf,subcaption}

\newcommand{\RES}{{\tt POWHEG BOX RES}\xspace}
\newcommand{\PWG}{{\tt POWHEG BOX}\xspace}
\newcommand{\PY}{{\tt Pythia8.3}\xspace}

\newcommand{\HW}{{\tt Herwig7.2}\xspace}
\newcommand{\mathd}{{\rm d}}

\definecolor{azure}{rgb}{0.0, 0.5, 1.0}

\newcommand\sss{\mathchoice%
{\displaystyle}%
{\scriptstyle}%
{\scriptscriptstyle}%
{\scriptscriptstyle}%
}

\newcommand{\kt}{k_{\sss\rm T}}

\def\beq{\begin{equation}}
\def\beqn{\begin{eqnarray}}
\def\eeq{\end{equation}}
\def\eeqn{\end{eqnarray}}

\def\lq{\left[} 
\def\rq{\right]}

\def\({\left(} 
\def\){\right)} 

\newcommand     \MSB            {\ifmmode {\overline{\rm MS}} \else
                                 $\overline{\rm MS}$\fi}
\newcommand\nlf{n_{\rm lf}}
\newcommand\as{\alpha_{\sss\rm S}}
\newcommand\muf{\mu_{\sss\rm F}}
\newcommand\mur{\mu_{\sss\rm R}}

\newcommand\tf{T_{\sss F}}

\newcommand{\mWT}{m_{\sss T W}}
\newcommand{\mW}{m_{\sss W}}
\newcommand{\mZ}{m_{\sss Z}}
\newcommand{\HT}{H_{\sss T}}
\newcommand{\MiNNLOPS}{{\sc\small MiNNLO\textsubscript{PS}}}

\preprint{CERN-TH-2023-064}

\title{NLO + parton-shower generator for $\boldsymbol{W c}$ production in
  the \RES{}}

\author[a]{Silvia Ferrario~Ravasio,}
\author[b]{Carlo Oleari}
\emailAdd{silvia.ferrario.ravasio@cern.ch}
\emailAdd{carlo.oleari@mib.infn.it}

\affiliation[a]{CERN, Theoretical Physics Department, CH-1211 Geneva 23, Switzerland}

\affiliation[b]{Universit\`a degli Studi di Milano\,-\,Bicocca, and INFN,
  Sezione di Milano\,-\,Bicocca, Piazza della Scienza 3, 20126 Milano, Italy} 

\abstract{
We present the implementation of a next-to-leading-order plus
parton-shower event generator for the hadronic production of a heavy
charm quark accompanied by a leptonically-decaying $W$ boson in the
\RES{} framework.  We consider both signatures, i.e.~$pp\to c\, \ell^-
\bar{\nu}_{\ell}$ and $pp\to \bar{c} \,\ell^+ {\nu}_{\ell}$, and we
include exactly off-shell and spin-correlation effects, as well as
off-diagonal Cabibbo-Kobayashi-Maskawa contributions.
We present particle-level results, obtained interfacing our code with the
\HW{} and \PY{} shower Monte Carlo event generators, including hadronization
and underlying-event effects, and compare them against the data collected by the CMS
Collaboration at $\sqrt{s}=13$~TeV.
}
\begin{document}

\maketitle

\section{Introduction}

The production of a $W$ boson in association with a charm quark~($c$), tagged
by the full reconstruction of the charmed $D$ hadron from the clear
experimental signature of its decay products, is one of the main probe of the
strange-quark content of the colliding particles at an hadron collider, such
as the LHC.
In fact, the leading contribution to $Wc$ production ($W^- c$ and $W^+
\bar{c}$) comes from the scattering of a strange quark ($s$ or $\bar{s}$) and
a gluon, from the two incoming hadrons, since the corresponding
Cabibbo-Kobayashi-Maskawa~(CKM)  matrix element is the dominant one for this
channel. The two leading-order~(LO) Feynman diagrams for $Wc$ production are
depicted in Fig.~\ref{fig:diagr-Born}.

As first illustrated in Ref.~\cite{Baur:1993zd}, this process can be used to put
constraints on the strange-quark content of the colliding hadrons,
i.e.~protons at the Large Hadron Collider~(LHC).  Several measurements of
$Wc$ production have already been performed at the LHC as it went through the
increasing energy upgrade from 7 to 13~TeV, by the ATLAS~\cite{ATLAS:2014jkm,
  ATLAS:2023ibp}, the CMS~\cite{CMS:2013wql, CMS:2021oxn, CMS:2018dxg} and
the LHCb Collaborations~\cite{LHCb:2015bwt}.

The next-to-leading order~(NLO) QCD cross
section for $Wc$ production has been known for a while, and studied both at
the Tevatron~\cite{Giele:1995kr} and at the LHC~\cite{Stirling:2012vh}.
More recently, the complete set of next-to-next-to-leading order~(NNLO) QCD
corrections to the dominant CKM-diagonal contribution have been computed in
Ref.~\cite{Czakon:2020coa} and further extended with full CKM dependence,
including the dominant NLO electro-weak corrections, in
Ref.~\cite{Czakon:2022khx}, in the massless-charm limit.

The first dedicated NLO QCD + parton shower~(PS) generator for $Wc$
production was done in Ref.~\cite{Bevilacqua:2021ovq}, within the {\tt
  PowHel} event generator, based on the POWHEG method~\cite{Nason:2004rx},
with the $W$ boson decaying leptonically, including charm-quark mass effects
in the hard-scattering matrix elements and a non-diagonal CKM
matrix.\footnote{Predictions for $Wc$ hadro-production, at NLO QCD accuracy,
matched to PS, according to the MC@NLO matching
framework~\cite{Frixione:2002ik}, can also be obtained using the {\tt
  MadGraph5\_aMC@NLO} framework~\cite{Alwall:2014hca}. In addition, one can
also use the authomatic interface between the \PWG and {\tt
  MadGraph5\_aMC@NLO}~\cite{Nason:2020lxx} to build the code, according to
the POWHEG matching framework.}

In this paper we describe the implementation of a NLO + parton-shower
generator for the hadro-production of a massive charm quark accompanied by a
leptonically-decaying $W$ boson in the \RES framework~\cite{Frixione:2007vw,
  Alioli:2010xd, Jezo:2015aia}.
We include exactly off-shell and spin-correlation effects, as well as
off-diagonal CKM contributions.

The \RES framework is able to deal with radiation off resonances. Although
for the case at hand of QCD corrections to a leptonically decaying $W$ boson,
the machinery of radiation off resonances is not necessary, since no QCD
radiation can be emitted from the leptons, we have implemented $Wc$
production in this framework for several reasons:
\begin{enumerate}
\item Better handling of final-state radiation from heavy quarks, as
  radiation collinear to heavy partons is dealt with an appropriate
  importance sampling.\footnote{The first treatment of radiation from
  heavy parton in the \PWG{} appeared in Ref.~\cite{Barze:2012tt}.}
  
\item The future inclusion of NLO electro-weak corrections is
  facilitated, since the process is already in the right framework
  to deal with photon radiation, not only from quarks, but also from
  the charged lepton in the $W$ decay, where the virtuality of the
  resonance has to be preserved during the generation of the hardest
  radiation performed by the \RES.

\item All the processes for which the NNLO+PS accuracy has been
  reached~\cite{Mazzitelli:2020jio, Mazzitelli:2021mmm, Zanoli:2021iyp,
    Gavardi:2022ixt} using the \MiNNLOPS{} formalism~\cite{Monni:2019whf,
    Monni:2020nks} are implemented in the \RES framework.  It would then be
  easier to implement also the NNLO QCD corrections to $Wc$ production in
  this framework too.  
\end{enumerate}
The paper is organized as follows. In Sect.~\ref{sec:proc} we give some
technical details about the process we are studying at LO and NLO, we
describe the change of scheme of our calculation and the default choice of
the renormalization and factorization scales that we have used. In addition
we give the numerical value of the input parameters we have used in our
simulation, and we present the validation of the fixed-order NLO results.  In
Sect.~\ref{sec:powhegmatching} we give details of the matching of the NLO
amplitudes with the POWHEG method, as implemented in the \RES{} framework. We
discuss the effects of the extra degrees of freedom offered by this matching
and their numerical impact.
In Sect.~\ref{sec:NLO+PS} we present a few interesting kinematic distributions
computed with events obtained by completing the first radiation emission
performed by the \RES{} code, with three showering models, implemented in
\HW{} and \PY, and we discuss the differences. We also present results with
full hadronization in place and in the presence of underlying events, and we
compare them with the experimental data collected by the CMS Collaboration at
13~TeV.

The \RES framework, together with the \texttt{Wc} generator, can be
downloaded at \url{http://powhegbox.mib.infn.it}.

\section{Contributing processes and technical details}
\label{sec:proc}

In this section we discuss the implementation of the processes $pp\to \ell^-
\bar{\nu}_{\ell} c$ and $pp\to \ell^+ \nu_{\ell} \bar{c}$ at NLO QCD
accuracy.
\begin{figure}[htb!]
\begin{center}
  \includegraphics[scale=0.7]{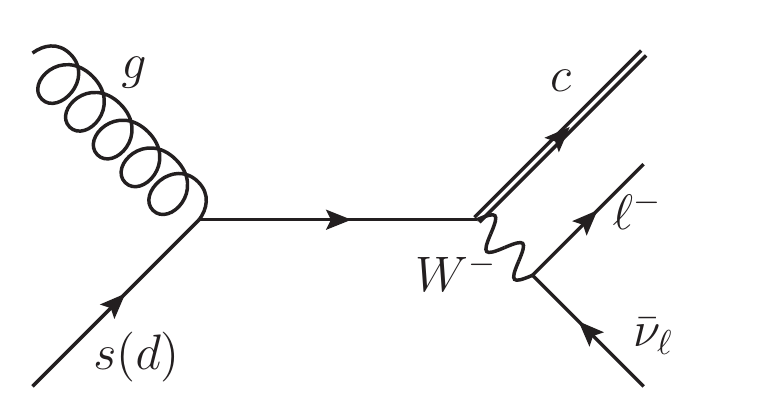}\hspace{1cm}
  \includegraphics[scale=0.7]{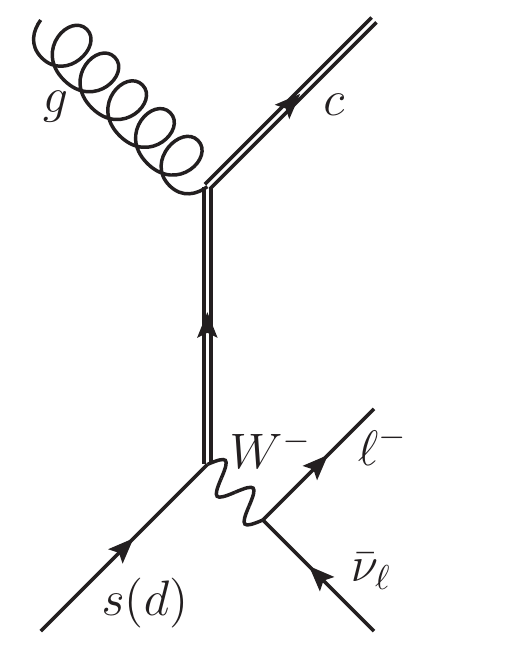}\\
  \vspace{0.2cm}
  \begin{minipage}[c]{0.31\textwidth}
    \subcaption{}\label{fig:born-s}
  \end{minipage}
  \hspace{1.8cm}
  \begin{minipage}[c]{0.2\textwidth}
    \subcaption{}\label{fig:born-t}
  \end{minipage}
\end{center}
\caption{Feynman diagrams contributing to the partonic process $g\, s(d)
  \rightarrow \ell^- \,\bar{\nu}_{\ell}\, c$ at LO, in the $s$ channel~(a), on
  the left, and $t$ channel~(b), on the right.  The heavy charm quark is
  represented as a double line.
\label{fig:diagr-Born}}
\end{figure}
We consider the $c$ quark to be massive, with
all the other lighter quarks and anti-quarks treated as massless.
At LO, the partonic subprocesses that contribute to $Wc$ production are
\begin{equation}
{\displaystyle
\begin{array}{l}
 g\, d \rightarrow  \ell^- \,\bar{\nu}_{\ell}\, c   \\
 g \,s \rightarrow  \ell^- \,\bar{\nu}_{\ell}\, c  \\
 g \,\bar{d} \rightarrow  \ell^+ \,\nu_{\ell}\, \bar{c}  \\
 g \,\bar{s} \rightarrow  \ell^+ \,\nu_{\ell}\, \bar{c}
 \label{eq:LOproc}
\end{array}
}
\end{equation}
where the channels involving a down (anti-)quark are Cabibbo-suppressed with
respect to those containing the strange (anti-)quark.  We are neglecting
contributions arising from a bottom quark in the initial state: these
contributions are suppressed both by the bottom parton distribution
function~(PDF), and also by the smallness of the CKM matrix element~$|V_{cb}|$. 

The tree-level Feynman diagrams contributing to the process $g \,s(d)
\rightarrow \ell^- \,\bar{\nu}_{\ell} \,c$ are represented in
Fig.~\ref{fig:diagr-Born}: this process can proceed either via an $s$-channel
exchange of a $s(d)$ quark (left panel), or via a $t$-channel exchange of a
$c$ quark (right panel).  The mass of the charm quark ensures that the latter
contribution is finite also when its transverse momentum is vanishing.

In the following we discuss only $W^-$ production, since a similar discussion
is also valid for $W^+$ production, after charge conjugation of the involved
subprocesses.

\begin{figure}[htb!]
\begin{center}
\includegraphics[scale=0.7]{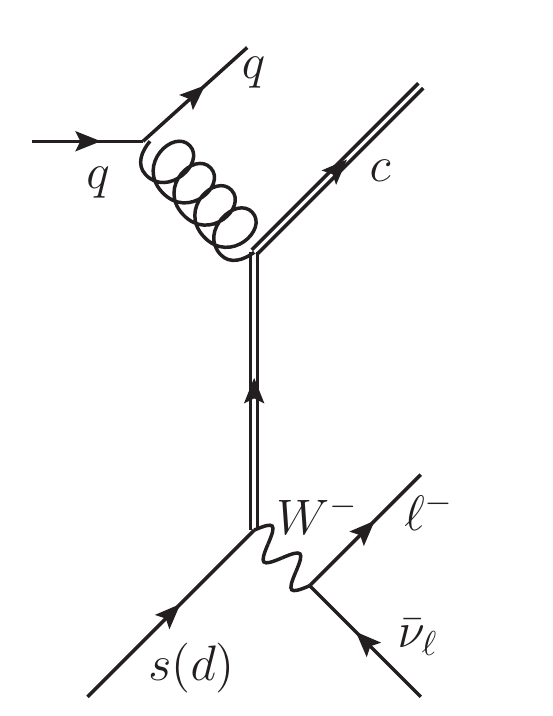} \hspace{1cm}
\includegraphics[scale=0.7]{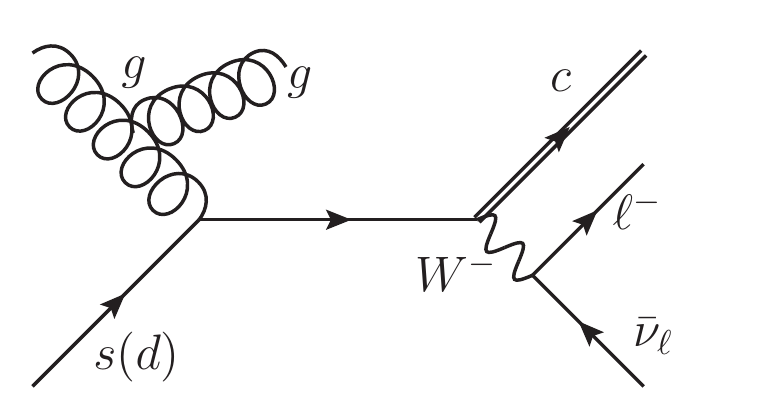}
\end{center}

  \hspace{2.5cm}
  \begin{minipage}[c]{0.2\textwidth}
    \subcaption{}\label{fig:realdiv1}
  \end{minipage}
  \hspace{2cm}
  \begin{minipage}[c]{0.31\textwidth}
    \subcaption{}\label{fig:realdiv2}
  \end{minipage}
\begin{center}
\includegraphics[scale=0.7]{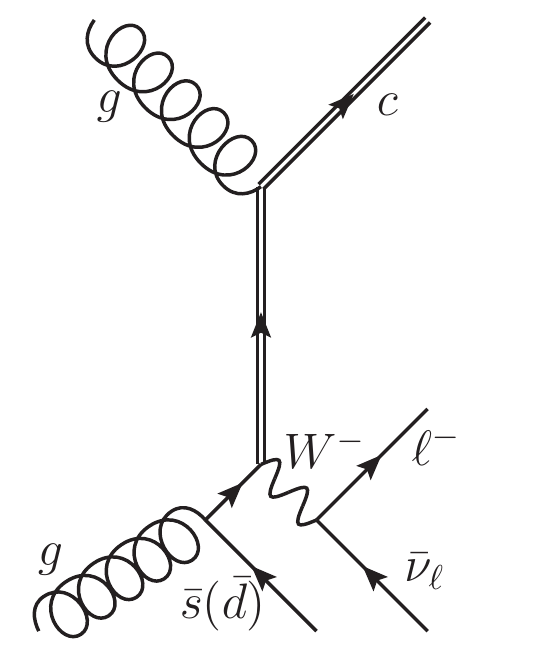} \hspace{1cm}
\includegraphics[scale=0.7]{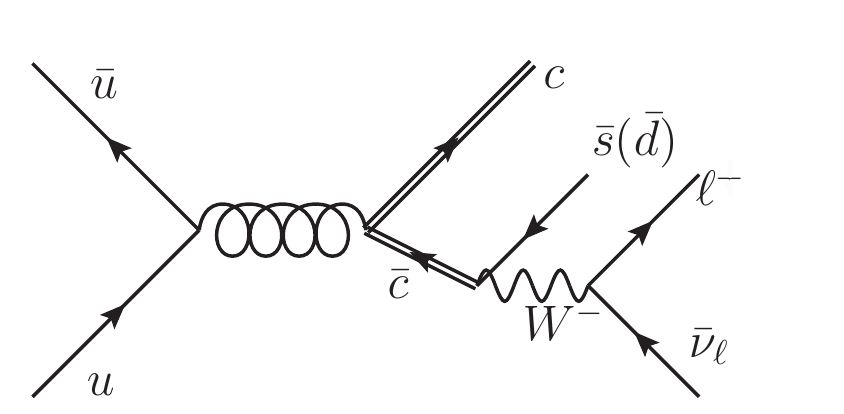} 
\end{center}

  \hspace{2.5cm}
  \begin{minipage}[c]{0.2\textwidth}
    \subcaption{}\label{fig:ggreal}
  \end{minipage}
  \hspace{2cm}
  \begin{minipage}[c]{0.31\textwidth}
    \subcaption{}\label{fig:qqreal}
  \end{minipage}
\caption{Examples of Feynman diagrams contributing to the real corrections
  for the process $pp \rightarrow \ell^- \bar{\nu}_{\ell} c\, j$, where
  $j=g,u,d,s,\bar{u},\bar{d},\bar{s}$. The last Feynman diagram~(d) is an example
  of ``regular'' contribution, i.e.~no singularities are associated with this
  process.
\label{fig:diagr-real}}
\end{figure}
Together with the QCD virtual corrections to the subprocesses in
Eq.~(\ref{eq:LOproc}), not depicted here, also the real corrections
contribute at the next-to-leading order.
Examples of Feynman diagrams contributing to the real corrections for the
process $pp \rightarrow \ell^- \bar{\nu}_{\ell} c\, j$ are shown in
Fig.~\ref{fig:diagr-real}.
We can separate the real-correction contributions into two classes:
\begin{enumerate}
  \item The corrections of the type $pp \rightarrow \ell^- \bar{\nu}_{\ell}
    \, c \, j$, with $j=g,u(\bar{u}),d (\bar{d}),s (\bar{s})$, that are singular
    in the limit of the unresolved jet

     \beqn
    \label{eq:real_proc}    
    && q(\bar{q}) \,s(d) \rightarrow  \ell^- \,\bar{\nu}_{\ell} \,c \,q(\bar{q}) \nonumber \\
    && g \,s(d) \rightarrow  \ell^- \,\bar{\nu}_{\ell}\, c\, g \\
    && g \,g \rightarrow  \ell^-\, \bar{\nu}_{\ell} \,c  \,\bar{s}(d)\,, \nonumber
    \eeqn
where $q=u,d,s$.  Examples of these contributions are depicted in
Fig.~\ref{fig:diagr-real}, diagrams~(a), (b) and~(c).
We will refer to these terms with $R$ in the following.

\item The corrections of the type
  \beq
  \label{eq:reg_proc}
  q \,\bar{q} \rightarrow \ell^-\, \bar{\nu}_{\ell} \,c\,  \bar{q}',
  \eeq
where $q=u,d,s$ and $q'= s, d$, which do not have any singularities
associated with the extra light-parton emission, and are called regular, in
the POWHEG jargon and will be indicated with $R_{\rm\sss reg}$. Again we do not
consider contributions with a final-state $b$ quark. An example of this type
of contributions is depicted in Fig.~\ref{fig:diagr-real}~(d). 

\end{enumerate}
Notice that the square of the diagrams contributing to the regular processes
(Fig.~\ref{fig:diagr-real}~(d)), as well as similar diagrams with two gluons
in the initial state (that interfere with the one in~(c)), suffer from
another potential source of divergence, due to the presence of the charm
anti-quark in the $s$ channel. In fact, when the invariant mass of the system
comprising the $\bar{s}(\bar{d})$ quark and the two leptons is equal to the
charm mass $m_c$, these diagrams are divergent, due to the fact that the
$c$-quark propagator goes on shell.
In order to avoid this singularity, we can impose a technical cut on the invariant mass
of the off-shell $W$ boson
\begin{equation}
  \label{eq:W_virt_cut}
 m_{\ell \nu}^2   \equiv (p_{\sss\ell}+p_{\sss\nu})^2 > m_c^2\,,
\end{equation}
in the theoretical simulation.  In order to make a closer contact to
experimentally accessible quantities, a more realistic cut is in general
imposed on the transverse mass of the $W$ system, $\mWT$, defined as
\begin{equation}
  \label{eq:mTw}
  \mWT^2= 2\, p_{\sss \perp \ell} \, p_{\sss \perp \nu} \(1-\cos \Delta \varphi\)\,,
\end{equation}
where $\Delta \varphi$ is the azimuthal separation between the transverse
momenta of the charged lepton and the missing energy of the neutrino.

We neglect contributions with two charmed quark in the final state, such as
$\bar{u} d \rightarrow \ell^- \,\bar{\nu}_{\ell} \,c \bar{c}$, coming from
the splitting $g\to c \bar{c}$, since these contributions are subtracted from
experimental data in studies aimed at extracting information on the
strange-quark parton distribution function, as illustrated in
Sect.~\ref{sec:anatomy}.

The Born and virtual matrix elements, as well as the Born colour- and
spin-correlated amplitudes, necessary to the \RES{} to automatically
implement the POWHEG formalism, have been computed
analytically.\footnote{In fact, the analytic calculation of these
amplitudes was part of the Master thesis of one of the authors.} The
one-loop amplitudes have been validated against the {\tt Gosam}
code~\cite{Cullen:2014yla, Cullen:2011ac}.  The matrix elements for
the real contributions were instead generated using {\tt
  Madgraph4}~\cite{Alwall:2007st} and its interface with the
\PWG{}~\cite{Campbell:2012am}.

\subsection[The decoupling and ${\overline{\rm MS}}$ schemes]
           {The decoupling and $\boldsymbol{\overline{\rm MS}}$ schemes}

When performing a fixed-order calculation with massive quarks, one can
define al least two consistent renormalization schemes that describe
the same physics: the usual \MSB{} scheme, where all flavours are
treated on equal footing, and a mixed scheme~\cite{Collins:1978wz},
called decoupling scheme~\cite{Appelquist:1974tg}, in which the $\nlf$
light flavours are subtracted in the \MSB{} scheme, while the
heavy-flavour loop is subtracted at zero momentum. In this scheme, the
heavy flavour decouples at low energies. This is the scheme we have
used in the renormalization of our analytic calculation, with
  $\nlf=3$.

If the decoupling scheme is chosen, then the strong coupling constant $\as$
runs with three light flavours, and the parton distribution functions should
not include the charm quark in the evolution.

To make contact with other results expressed in terms of the $\MSB$
strong coupling constant, running with five light flavours, and with
PDFs with five flavours, we prefer to
change our renormalization scheme and to switch to the $\MSB$ one.
The procedure for such a switch is well known, and was discussed in
Ref.~\cite{Cacciari:1998it}. This procedure, applied to the $b$ quark, is
also illustrated in Appendix~A of Ref.~\cite{Luisoni:2015mpa}.

For the case under study, since the LO process contains only one power of
$\as$, if we want to express our calculation in terms of a coupling constant
defined in the \MSB{} scheme with five active flavours, we need to add the
following contribution to the cross section
\begin{equation}
   \label{eq:delta_as}
\Delta V_{\as}(\mur; m_b, m_c) = - \frac{1}{3} \tf \frac{\as}{\pi}
\left( \log \frac{\mur^2}{m_c^2}+\log \frac{\mur^2}{m_b^2}\right) B\, ,
\end{equation}
where $B$ is the Born cross section, $m_b$ the bottom mass and $\mur$ is the
renormalization scale.  A
corresponding modification has to be applied if we want to employ PDFs with
five active flavours.  The gluon PDF, which appears in the LO cross
section, induces a correction
\begin{equation}
  \label{eq:delta_g}
\Delta V_{g}(\muf; m_b, m_c) = +  \frac{1}{3} \tf \frac{\as}{\pi} \left(
\log \frac{\muf^2}{m_c^2}+\log \frac{\muf^2}{m_b^2}\right) B\,,
\end{equation}
where $\muf$ is the factorization scale.
The quark PDF receives only corrections that starts at order $\as^2$, and
can thus be neglected.
By adding Eqs.~(\ref{eq:delta_as}) and~(\ref{eq:delta_g}), we get the
conversion factor for the two schemes
\begin{equation}
  \label{eq:deltaV}
\Delta V (\muf,\mur)= \Delta V_{\as} + \Delta V_{g} = \frac{2}{3}\, \tf
\frac{\as}{\pi} \, B \log \frac{\muf^2}{\mur^2}\,, 
\end{equation}
that turns out to be independent from the bottom and charm masses, and
different from zero only if the renormalization and factorization scales are
different.

\subsection{Renormalization and factorization scale settings}
\label{sec:ren_fac_scales}
In the \PWG{} there is the option to use a different renormalization
and factorization scale when computing the real contribution (and/or
the subtraction terms, i.e.~their soft and collinear limits) or the
Born and the virtual ones. The only requirement is that the scale used
in the evaluation of the real contributions (and possibly of the
subtraction terms) must tend to the scale used in the Born amplitude,
when the emitted parton gets unresolved.

The default central scale used in our simulation in the evaluation of the
Born, virtual, real and subtraction terms is
\begin{equation}
\label{eq:mu}
\mu = \frac{\HT}{2}
\end{equation}
with
\begin{equation}
  \label{eq:H_T}
\HT =
\sqrt{|\vec{p}_{\perp c}|^2+m_c^2}+\sqrt{ |\vec{p}_{\perp \ell} +
  \vec{p}_{\perp \nu}|^2 + (p_\ell+p_\nu)^2}\,,
\end{equation}
evaluated with the kinematics of the underlying-Born configuration.
The regular real term $R_{\rm\sss reg}$ has no underlying Born configuration,
and the scale $\HT$ is computed using the kinematics of the real emission
contribution, adding also the transverse momentum of the massless emitted parton.
\begin{equation}
  \label{eq:HTj}
 \HT = \sqrt{|\vec{p}_{\perp c}|^2+m_c^2}+\sqrt{ |\vec{p}_{\perp \ell} +
   \vec{p}_{\perp \nu}|^2 + (p_\ell+p_\nu)^2} + \left|\vec{p}_{\perp j}\right|\,.
\end{equation}
Other options contemplate the use of the scale~(\ref{eq:HTj}) also in the
real term and/or in the subtraction terms,\footnote{This option can be chosen
adding the flags {\tt btlscalereal 1} and/or {\tt btlscalect 1} in the
input card.} or a fixed scale,\footnote{Set {\tt runningscales 0} in the
input card.} such as $\mu=\mW$.

The renormalization and factorization scales, $\mur$ and $\muf$, are defined
multiplying $\mu$ by the factors $\xi_{\sss R}$ and $\xi_{\sss F}$
respectively.  The uncertainty due to missing higher order corrections is
estimated considering the 7-point scale variations
\begin{equation}
  \label{eq:7-point}
(\xi_{\sss R},\xi_{\sss F}) = (1,1), \, (1,2), \, (1,0.5), \, (2,1), \,
  (0.5,1), \, (2,2), \,  (0.5,0.5). 
\end{equation}
We employ the NNPDF3.1\_nlo\_pdfas~\cite{Buckley:2014ana, NNPDF:2017mvq} PDF
set, which is also used to determine the running of the strong
coupling.\footnote{This option can be selected by adding the flag
{\tt alphas\_from\_pdf 1} in the input card. Otherwise the running of
$\as$ is computed at two loops internally by the \RES{}.}

\subsection{Numerical inputs}
\label{sec:params}
We have simulated $Wc$ production in $pp$ collisions at $\sqrt{s}=13$~TeV.
The input parameters we have used are
\begin{align}
  \label{eq:phys_values}
&\mW =  80.385~{\rm GeV}\phantom{^{-2} \times 10^{-5}}  \qquad \Gamma_{\sss
    W} = 2.085~{\rm GeV}
  \nonumber\\
&\mZ =  91.1876~{\rm GeV}\phantom{^{-2} \times 10^{-5}} \qquad \Gamma_{\sss Z} = 2.4952~{\rm GeV}\\
  &G_\mu =  1.16639 \times 10^{-5}~{\rm GeV}^{-2}\, \qquad m_c = 1.5~{\rm GeV}.
  \nonumber
\end{align} 
The electromagnetic coupling is obtained in the $G_\mu$ scheme
\begin{equation}
\alpha_{\rm em} = \frac{\sqrt{2}}{\pi}
G_{\mu}\mW^2\left(1-\frac{\mW^2}{m_{\sss Z}^2}\right),
\end{equation}
and the values of relevant CKM matrix elements are
\begin{equation}
|V_{cd}| = 0.22438\,, \qquad |V_{cs}| = 0.97356\,.
\end{equation}
In our calculation we have neglected the mass of the leptons.  Only at the
stage of event generation a leptonic mass is assigned to the charged lepton
once the full kinematics of the event is generated.\footnote{The user can
also choose in which leptonic family/families the $W$ boson decays, by selecting the
value of the flag {\tt vdecaymode} in the input card.}
The assignment of the mass is done through a reshuffling procedure that
preserves the total energy and momentum, and the virtuality of the resonances.
The values of the lepton masses we have used are given by
\begin{equation}
m_e = 0.511\times 10^{-3}~{\rm GeV}, \qquad m_\mu = 0.1057~{\rm GeV},
\qquad m_\tau = 1.777~{\rm GeV}.
\end{equation}
Furthermore, following the discussion that precedes
Eq.~(\ref{eq:W_virt_cut}), we always apply a technical loose cut, at the
generation stage, on the $W$-boson virtuality
\begin{equation}
  \label{eq:tech_cut}
  m_{\ell \nu}  > 2~{\rm GeV}\,.
\end{equation}

\subsection{Fixed-order validation}

\label{sec:MCFMvalidation}
\begin{figure}[htb]
\begin{subfigure}{0.5\textwidth}
  \includegraphics[height=0.31\textheight,page=9]{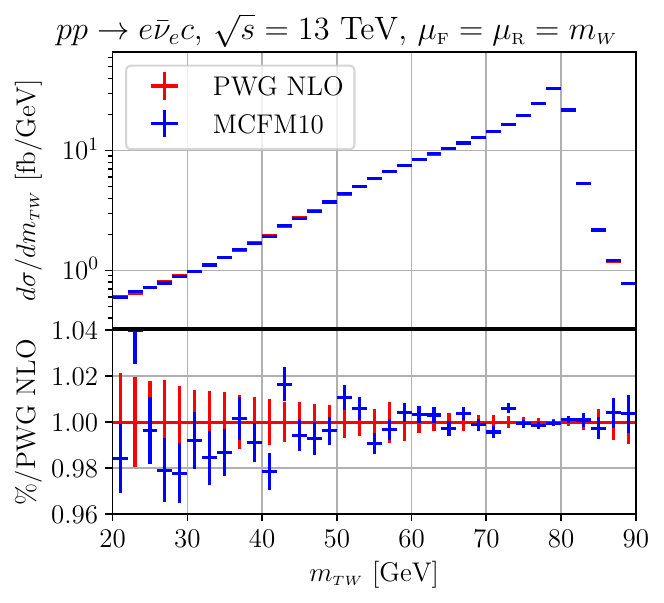}
\end{subfigure}%
\begin{subfigure}{0.5\textwidth}
  \includegraphics[height=0.31\textheight,page=11]{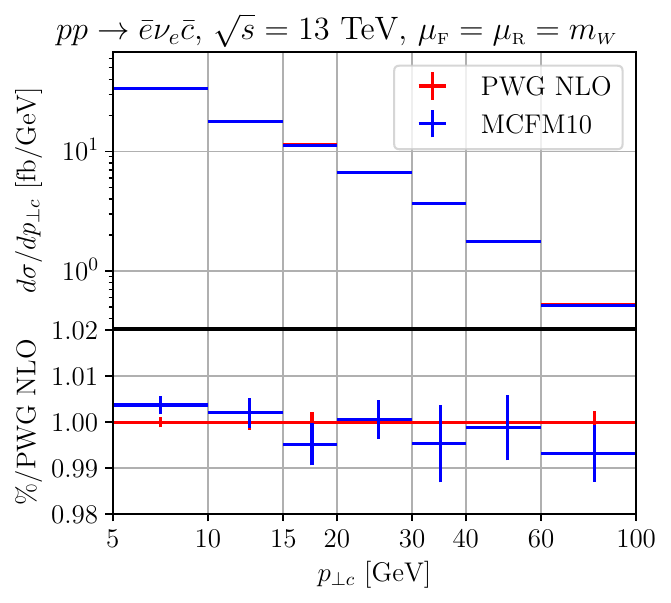}
\end{subfigure}
\caption{Differential distribution for the transverse mass of the
  reconstructed $W^-$ boson for the processes $pp\to e^- \bar{\nu}_e
  c$~(left) and the transverse momentum of the heavy anti-charm quark for the
  processes $pp \to e^+ \nu_e \bar{c}$~(right) at NLO accuracy. The input
  parameters are discussed in Sect.~\ref{sec:params}. The ratio between
  predictions obtained with {\tt MCFM}~(blue) and the \RES~(red) is also
  shown. The statistical error bars from the Monte Carlo integrator are also
  shown. 
\label{fig:MCFM}}
\end{figure}

In this section we compare our fixed-order results with those obtained with
the {\tt MCFM} code~\cite{Campbell:2019dru, Campbell:2015qma}.
In order to perform the comparison we employ a fixed
renormalization and factorization scale set to
\begin{equation}
\muf = \mur = \mW\,.
\end{equation}
As we are setting $\muf=\mur$, the term $\Delta {\rm V}$ defined in
Eq.~\eqref{eq:deltaV} vanishes and there is no correction factor.
We define the fiducial cross section imposing the following cuts on the
transverse momentum and rapidity of the charged lepton, as well on the
transverse momentum of the   charm quark
\begin{equation}
  \label{eq:CMScuts}
|\vec{p}_{\perp \ell}|>26~{\rm GeV} , \qquad |y_\ell|<2.4\, , \qquad
|\vec{p}_{\perp c}| > 5~{\rm GeV}.
\end{equation}
These cuts corresponds to the ones employed by the CMS Collaboration to
define the inclusive cross section in the analysis described in
Ref.~\cite{CMS:2018dxg} at $\sqrt{s}=13$~TeV. 

As discussed in Sect.~\ref{sec:proc}, around Eqs.~(\ref{eq:W_virt_cut})
and~(\ref{eq:mTw}), in order to avoid the charm-propagator singularity in
diagrams like the one in Fig.~\ref{fig:diagr-real}~(d),
we also implement a cut on the transverse mass of the $W$ system
\begin{equation}
\mWT > 20~{\rm GeV}.
\label{eq:mTwcut}
\end{equation}
In Table~\ref{tab:xsecMCFM}
\begin{table}[htb]
  \begin{center}
\begin{tabular}{|c|c|c|c|c|}
\hline
$\sigma$ [fb] & $e^- \bar{\nu}_e c$ @ LO & $e^+ \nu_e \bar{c}$ @ LO &
$e^- \bar{\nu}_e c$ @ NLO & $e^+ \nu_e \bar{c}$ @ NLO \\
\hline
{\tt PWG RES}          & 352.7(1) & 341.1(1) & 497.2(2) & 480.8(2) \\
{\tt MCFM 10} & 352.9(2) & 341.2(2) & 497.0(2) & 481.0(3) \\
Ratio            & 1.0005(5) & 1.0002(5) & 0.9996(7) & 1.0004(7) \\
\hline
\end{tabular}
\caption{Fiducial cross section obtained with {\tt MCFM} and our
  generator ({\tt PWG RES}) for the processes $pp\to e^- \bar{\nu}_e
  c$ and $e^+ \nu_e \bar{c}$ at LO (left columns) and NLO (right
  columns) at $\sqrt{s}=13$~TeV within the fiducial cuts of
  Eqs.~\eqref{eq:CMScuts} and~\eqref{eq:mTwcut}, with $\muf=\mur=\mW$.
  The input parameters are discussed in Sect.~\ref{sec:params}.  In
  the last row we show the ratio between the {\tt MCFM} and the \RES{}
  predictions.  }
\label{tab:xsecMCFM}
  \end{center}
\end{table}
we show the integrated cross section at LO and NLO obtained with our
code and with {\tt MCFM}. In all cases a sub-permille agreement is
found.  Good agreement is found also at the differential level, as it
can be seen in Fig.~\ref{fig:MCFM}, where we illustrated the
transverse mass of the reconstructed $W$ boson~(left panel) and the
transverse momentum of the charmed quark~(right panel).

\section{Details of the POWHEG matching}
\label{sec:powhegmatching}

In this section we briefly recall the expression for the differential
cross section generated with the POWHEG method~\cite{Nason:2004rx} for the
process under consideration, focusing for simplicity only on the $pp\to \ell
\bar{\nu}_{\ell} c$ signature, in order to discuss further sources of 
theoretical uncertainty intrinsic to the POWHEG method. 

The NLO cross section reads
\begin{align}
  \label{eq:sNLO}
\mathd\sigma_{\rm\sss NLO} = B(\Phi_{3};\mur,\muf)\, \mathd\Phi_{3} +
\frac{\as(\mur)}{2\pi} & \lq V(\Phi_{3};\mur,\muf) \, \mathd\Phi_{3} +
     R(\Phi_{3+1}; \mur,\muf) \, \mathd\Phi_{3+1} \right.
     \nonumber\\
     &     \left. {} + R_{\rm \sss reg} (\Phi_{3+1};\mur,\muf)\,  \mathd\Phi_{3+1}\rq,
\end{align}
where $B$, $V$, $R$ and $R_{\rm \sss reg}$ are the Born, virtual, real
and regular real matrix elements, multiplied by the appropriate PDF
and flux factors. As stated in Sect.~\ref{sec:proc}, we indicate with
$R_{\rm \sss reg}$ the terms arising from the flavour configurations
of Eq.~\eqref{eq:reg_proc}, which do not contain any QCD singularity,
while the processes in Eq.~(\ref{eq:real_proc}) are designated with
$R$.  The Born and virtual cross sections are evaluated using a
three-body phase space $\Phi_3$, while, in the real contributions, an
additional light parton is present.  The matrix elements depend on the
$\mur$ scale, as the LO starts with one power of the strong coupling
constant $\as$, while the $\muf$ dependence arises from the
incoming-parton PDFs.  To build the POWHEG cross section, the real
term $R$ is split into a singular $R_{ \sss s}$ and a finite
contribution $R_{\sss f}$, and the $\bar{B}$ term is defined as
\begin{equation}
\label{eq:Bbar}
\bar{B}(\Phi_3;\muf,\mur) = B(\Phi_{3};\mur,\muf) +
\frac{\as(\mur)}{2\pi}\left[ V(\Phi_{3};\mur,\muf) +\int
  d \Phi_1 \,  R_{\sss s} (\Phi_{3+1};\mur,\muf)\right],
\end{equation}
where $\Phi_1$ is the radiation phase space, defined in terms of 
three radiation variables.
The probability of not having any resolved emissions up to a scale
$\kt^\prime$ is given by
\begin{equation}
\Delta_{\sss s}(\kt^\prime; \Phi_3) = \exp \left[- \int \mathd\Phi_1 \,
  \theta \!\lq \kt(\Phi_1)-\kt^\prime \rq \frac{\as(\kt)}{2\pi} \,
  \frac{{ R}_{\sss s}(\Phi_{3+1};\kt,\kt)}{{ B}(\Phi_3; \kt,
    \kt)}\right], 
\label{eq:sudakov}
\end{equation}
where the factorization and the renormalization scales are evaluated at the
scale $\kt$, which corresponds to the transverse momentum of the light
parton, if it is collinear to an initial-state parton, while, for gluon
emissions from the final-state charm quark, it is given by 
  \begin{equation}
    \kt^2 = 2 \frac{E_g}{E_c} \(p_g \cdot p_c\),
  \end{equation}
where $p_g$ and $p_c$ are the gluon and charm four-momenta and $E_g$ and
$E_c$ their energies in the partonic center-of-mass frame.

The POWHEG cross section finally
reads
\begin{align}
\label{eq:sPOWHEG}  
\mathd\sigma_{{\rm\sss PWG}} = &\bar{B}(\Phi_3;\muf,\mur) \, \mathd\Phi_3
\left[\frac{\as(\kt)}{2\pi}
  \frac{R_{\sss s}(\Phi_{3+1};\kt,\kt)}{{ B}(\Phi_3; \kt,
    \kt)} \, \Delta_{\sss s}(\kt;\Phi_3) \,\mathd\Phi_1  + \Delta_{\sss s}(\kt^{\sss \rm min}; \Phi_3)\right]
\nonumber \\
&+\frac{\as(\mur)}{2\pi} R_{\sss f}(\Phi_{3+1};\mur,\muf) \, \mathd\Phi_{3+1}
+\frac{\as(\mur)}{2\pi} R_{\rm\sss reg}(\Phi_{3+1};\mur,\muf) \, \mathd\Phi_{3+1}\,,
\end{align}
where $\kt^{\sss \rm min}$ is an infrared cutoff.

\begin{figure}[htb]
\centering
\includegraphics[width=0.5\textwidth]{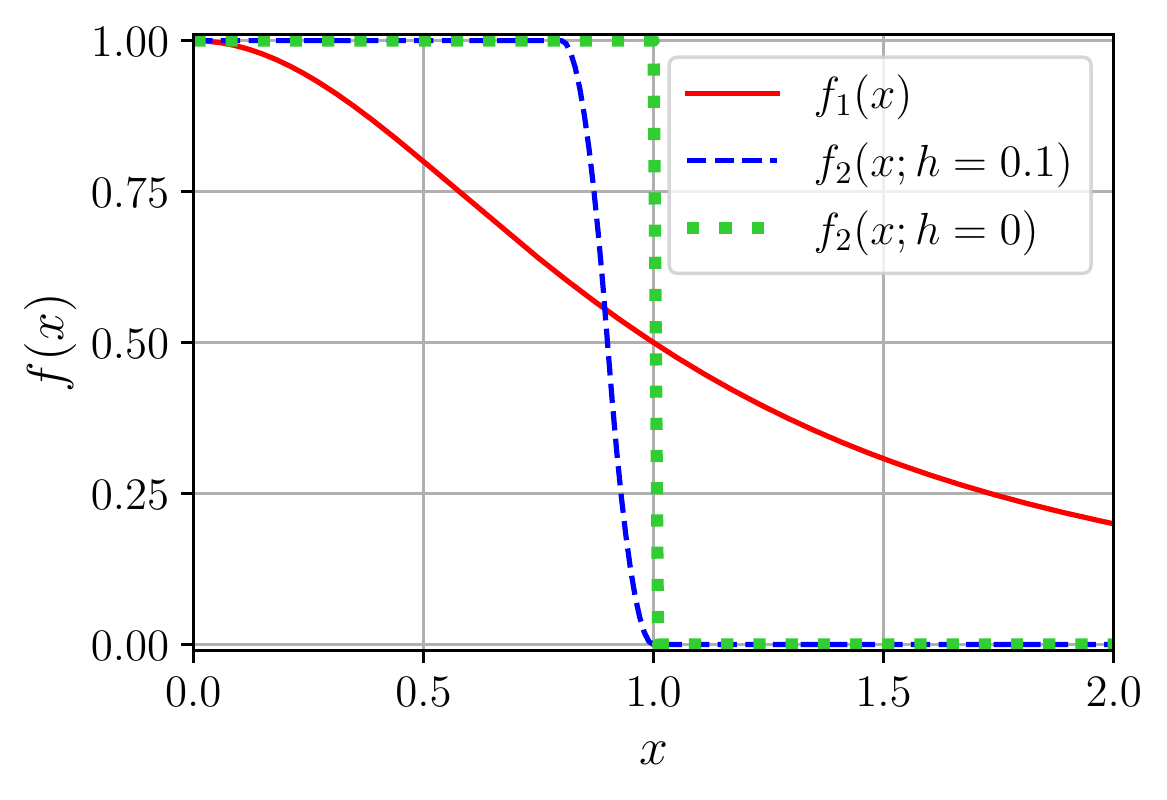}
\caption{Comparison of three damping functions used to separate the real
  contribution into a singular and a finite part, as detailed in
  Sect.~\ref{sec:powhegmatching}.
\label{fig:hdamp}}
\end{figure}
There are several degrees of freedom in the implementation of
Eq.~\eqref{eq:sPOWHEG}. One of them is in the choice of the renormalization
and factorization scales $\mur$ and $\muf$ in the $R_{\sss f}$ and
$R_{\rm\sss reg}$ terms, as well as in the term $R_{\sss s}$ appearing in
$\bar{B}$ in Eq.~\eqref{eq:Bbar}, as discussed in Sect.~\ref{sec:ren_fac_scales}.
Another one is in the choice of how to separate $R$ into a singular part
$R_{\sss s}$ and a finite one $R_{\sss f}$. Without loss of generality, one
can separate the two contributions introducing what is called a ``damping
function'', $f(\kt)$, that satisfies
 \begin{equation}
 0 \leq f(\kt) \leq 1\,,\qquad \lim_{\kt \to 0} f(\kt)=1
 \end{equation}
and define
\begin{equation}
  R_{\sss s} = f(\kt) \, R, \qquad
  R_{\sss f} = \lq 1-f(\kt) \rq R\,.
\label{eq:Rsf}
\end{equation}
For the process we are studying, we have considered two different forms for
$f$
 \begin{align}
 f_1(x) = &\hspace{0.2cm} \frac{1}{1+x^2} \label{eq:f1}\\
 f_2(x;h) = &
 \begin{cases}
  1 \quad &\mbox{ for } x\le1-2h\\
   {\displaystyle1-\frac{(1-2h-x)^2}{2h^2}} &\mbox{ for } 1-2h<x\le1-h\\[2mm]
  {\displaystyle\frac{(1-x)^2}{2h^2}} &\mbox{ for } 1-h<x\le1 \\[2mm]
  0 &\mbox{ for } x>1,
 \end{cases}
 \label{eq:f2}
 \end{align}
with 
\begin{equation}
x = \frac{1}{\mu}\sqrt{\kt^2- m_c^2} \,\, \theta\!\(\kt - m_c\),
\end{equation}
where $\mu$ is defined in Eq.~\eqref{eq:mu}, and $0\leq h \leq 1/2$ is
a free parameter.  The value of $x$ is set to 0 if $\kt<m_c$, so that
no finite contribution is present with $\kt$ below the charm
mass.\footnote{In the \RES, no damping factor is applied when the
emitter is massive.} The functional form $f_2$ in Eq.~\eqref{eq:f2},
with $h=0.3$, is the default damping function in the \HW{} event
generator~\cite{Bellm:2016rhh}, while $f_1$ in Eq.~\eqref{eq:f1}
corresponds to the default option of the \RES{}.
In Fig.~\ref{fig:hdamp} we provide a graphical representation of $f_1$,
$f_2(h=0.1)$ and $f_2(h=0)$.\footnote{Furthermore, the code has by default
the {\tt Bornzerodamp} flag activated. This provides a way to deal with
kinematic configurations where the real contribution is too large with respect to its
soft and collinear limits, signalling that the kinemtics of the underlying
Born gives rise to a Born amplitude that is particularly small, i.e.~close to
zero. See Ref.~\cite{Alioli:2010xd} for more details.}

\begin{table}[tb]
\centering
\begin{tabular}{|c|c|c|c|}
\hline
acronym & scales in $R$  & damping function $f$\\
\hline
B${}_0$ & underlying-Born kinematics      & 1 \\
\hline
B${}_1$ & underlying-Born kinematics      & $f_1$  \\
\hline
B${}_2$ & underlying-Born kinematics      & $f_2(h=0.1)$  \\
\hline
R${}_0$ & real kinematics      & 1 \\
\hline
\end{tabular}
\caption{Setups for the implementation of the POWHEG cross section given in
  Eq.~\eqref{eq:sPOWHEG}. The second column denotes the kinematic
  configuration used to evaluate the renormalization and factorization scales
  in the computation of the real and subtraction terms, while the last column
  shows which profile function is used to evaluate Eq.~\eqref{eq:Rsf}. The
  functional forms of $f_1$ and $f_2$ are given in Eqs.~\eqref{eq:f1}
  and~\eqref{eq:f2} respectively.}
\label{tab:hdamp}
\end{table}
In the following we investigate the dependence of several inclusive (with
respect to the QCD radiation) observables, such as the leptonic observables
and the rapidity and the transverse momentum of the jet containing the charm
quark.

We reconstruct jets using the Fastjet~\cite{Cacciari:2011ma} implementation
of the anti-$\kt$ algorithm~\cite{Cacciari:2008gp} with $R=0.5$.  To this
aim, we have considered the setups listed in Table~\ref{tab:hdamp} to
compute the POWHEG cross section given in Eq.~\eqref{eq:sPOWHEG}.  The
central value of the renormalization and factorization scales is set to
$\HT/2$ (see Eq.~(\ref{eq:mu})).

In the following figures we will compare some kinematic distributions
computed with six different setups: two fixed-order NLO results, one computed using
the underlying-Born kinematics (NLO$_{\rm B}$) for the renormalization and
factorization scale in the real contributions, and the other one using
the full real-event kinematics, indicated with NLO$_{\rm R}$.  The other four
distributions are computed analysing the POWHEG-generated events at the ``Les
Houches''~(LHE) level, i.e.~events with the first hardest radiation generated
according to Eq.~(\ref{eq:sPOWHEG}), adopting the setup scales and damping
functions of Table~\ref{tab:hdamp}. 
For inclusive observables, the LHE results should agree with the
fixed-order NLO ones up to NNLO terms.

\begin{figure}[htb!]
  \begin{subfigure}{0.515\textwidth}
    \includegraphics[width=\textwidth,page=12]{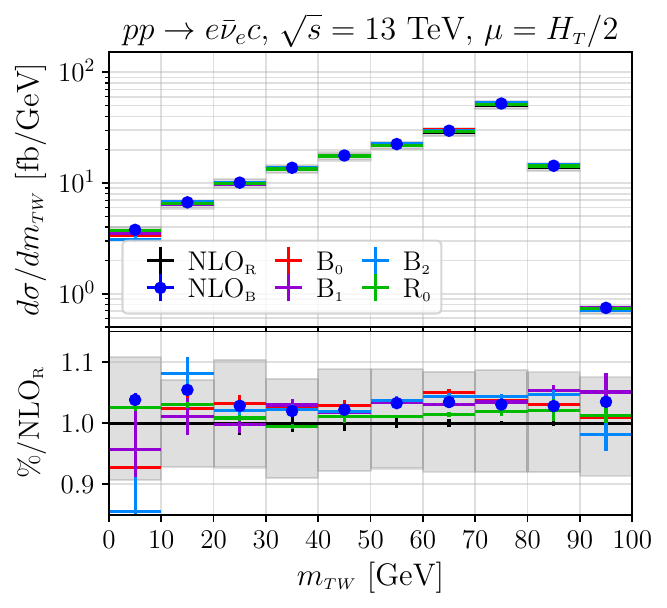}
  \end{subfigure}
  \hspace{-2.mm}
    \begin{subfigure}{0.515\textwidth}
     \includegraphics[width=\textwidth,page=6]{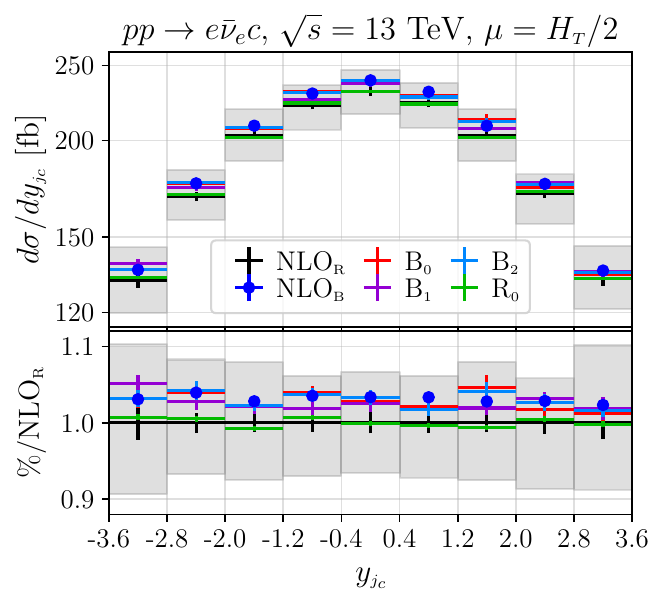}
\end{subfigure}
\caption{The transverse mass distribution of the reconstructed $W$
  boson~(left) and the rapidity of the charmed jet~(right) at NLO
  (NLO$_{{\rm B}}$ and NLO$_{{\rm R}}$) and at the Les Houches event
  level, for the choices of the scales and damping functions listed in
  Table~\ref{tab:hdamp}. The grey bands are obtained performing the
  7-point scale variations in Eq.~(\ref{eq:7-point}) of the NLO$_{{\rm
      R}}$ results. The statistical errors from the integration
  procedure are also shown.
\label{fig:no-pt-incl}}
\end{figure}
In Fig.~\ref{fig:no-pt-incl} we plot the transverse mass of the $W$ boson
(left panel) and the rapidity of the charmed jet (right panel).  The
NLO$_{{\rm B}}$ results (blue line) are roughly 3\% higher than the
NLO$_{{\rm R}}$ ones (black line), although always contained in the grey
scale-uncertainty band of NLO$_{{\rm R}}$, computed performing the 7-point
scale variations in Eq.~(\ref{eq:7-point}).  Furthermore, all the
distributions at the LHE level agree with the corresponding fixed order
calculation, within the scale-variation band.

\begin{figure}[htb]
  \begin{subfigure}{0.515\textwidth}
    \includegraphics[width=\textwidth,page=2]{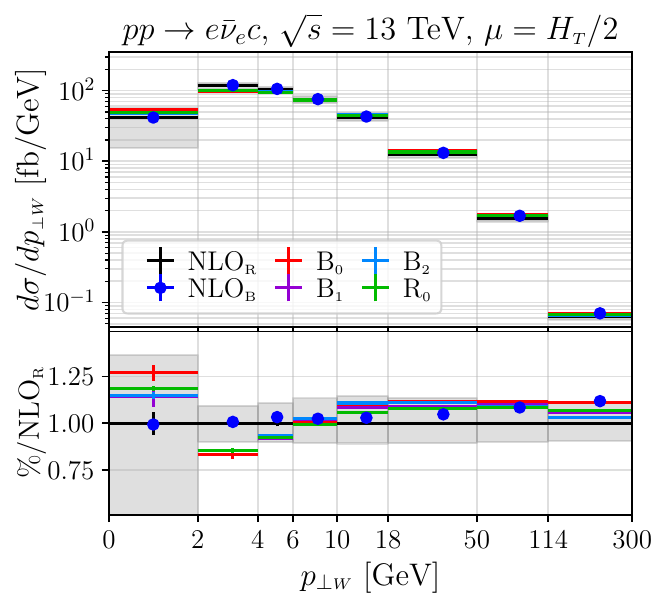}
  \end{subfigure}
  \hspace{-2.mm}
    \begin{subfigure}{0.515\textwidth}
     \includegraphics[width=\textwidth,page=4]{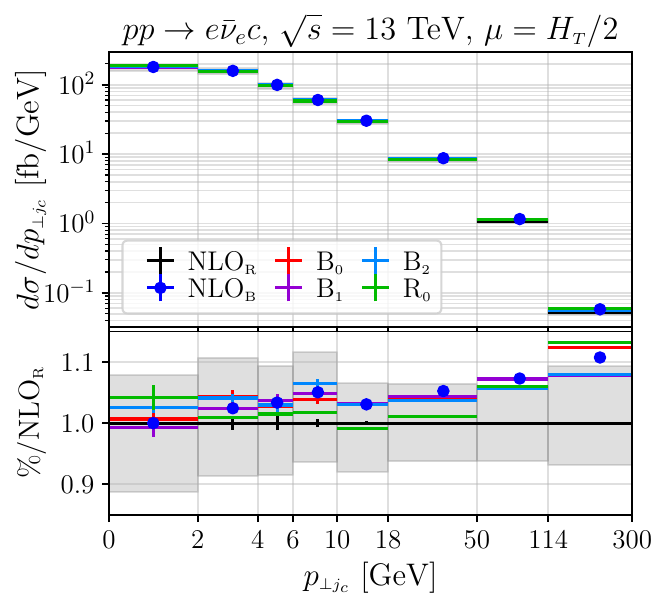}
\end{subfigure}
\caption{Same as Fig.~\ref{fig:no-pt-incl} but for the transverse momentum of
  the reconstructed $W$ boson~(left) and of the charmed jet~(right). 
\label{fig:pt-incl}}
\end{figure}
The transverse momentum of the $W$ boson, $p_{\sss \perp W}$, and of the
charmed jet, $p_{\sss \perp j_c}$, are illustrated on the left and right panel
of Fig.~\ref{fig:pt-incl}, respectively. Comparing the two NLO predictions
computed with different scales, one can see that the use of the
underlying-Born scales favours a harder spectrum, although the result always
lies within the scale-variation band of the NLO$_{\rm R}$ one, except for
the very last bin. This is due to the fact that, for large $c$-jet transverse
momentum, it is quite likely to produce a $c \to c g$ splitting that enhances
the value of $\HT$. As such, the underlying-Born kinematics yields a lower
scale and hence a larger value for $\as$, that enhances the cross section.
The same behaviour can be seen also at the LHE level.

\begin{figure}[htb]
\includegraphics[width=0.515\textwidth,page=9]{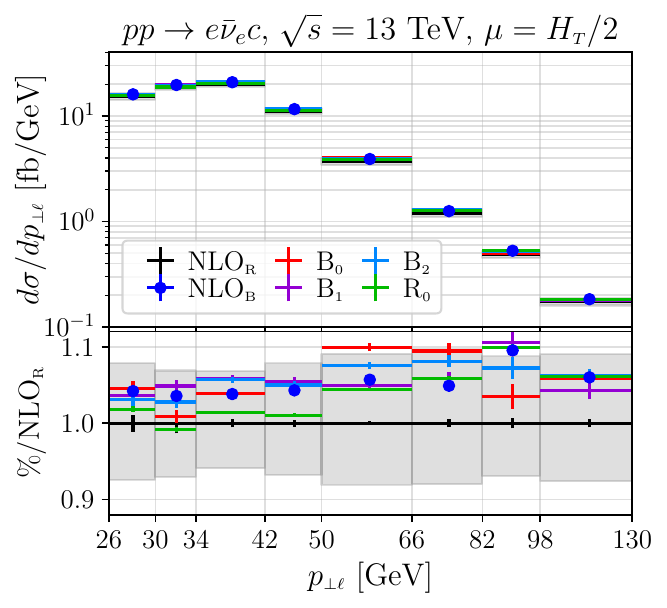}\hspace{-2.mm}
\includegraphics[width=0.515\textwidth,page=11]{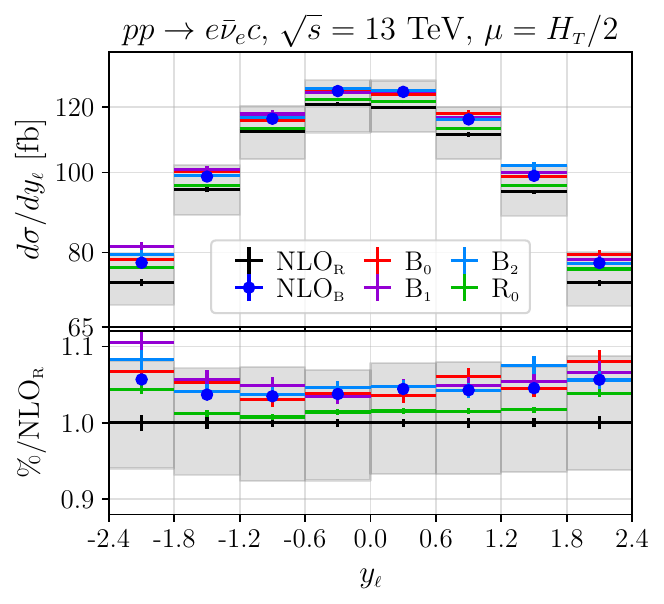}
\caption{Same as Fig.~\ref{fig:no-pt-incl} but
  for the transverse momentum~(left) and rapidity~(right) of the charged
  lepton, with the cuts in Eq.~\eqref{eq:CMScuts} and~\eqref{eq:mTwcut} in place.
\label{fig:lept-CMS}}
\end{figure}
In the experimental analyses, such as the one in Ref.~\cite{CMS:2018dxg}, an
important role is played by the measurement of the transverse momentum and
rapidity of the charged lepton. In Fig.~\ref{fig:lept-CMS} we then plot these
quantities, after applying the cuts of Eq.~(\ref{eq:CMScuts}), that are the
same as those used by the CMS Collaboration, as well as a cut on $\mWT$,
defined in Eq.~\eqref{eq:mTwcut}.  As far as the fixed-order NLO results is
concerned, the NLO$_{\rm B}$ curves lie a few percent above the NLO$_{\rm R}$
one, both for the transverse momentum and the rapidity. This is again due to
the higher value of $\as$ when the renormalization scale is evaluated with
the underlying-Born kinematics.  Differences up to 10\% can also be seen
among the LHE curves, but, in general, all these curves lie within the grey
uncertainty band of the NLO$_{\rm R}$ result, so that they are consistent
among each other.

We thus conclude that no significant differences are found among the several
damping factors, particularly when realistic analysis cuts are applied, and
that the main theoretical uncertainties arise from scale
variations.\footnote{In this analysis we are neglecting the dependence on the
PDF set. This topic is addressed in detail by the authors of
Refs.~\cite{Bevilacqua:2021ovq, Czakon:2022khx}, which considered several
sets, where the charm PDF is intrinsic or dynamically generated. The outcome
they found depends on the PDF set used in the simulation: for some sets, the
PDF-variation band is smaller than the scale-variation one, for others it is
comparable or slightly larger.}

\section{NLO + parton-shower results}
\label{sec:NLO+PS}
In this section, we present full results after the completion of the
POWHEG shower performed by general-purpose Monte Carlo~(MC) programs,
such as \HW~\cite{Bahr:2008pv, Bellm:2019zci} and
\PY~\cite{Sjostrand:2006za, Sjostrand:2014zea}, including
hadronization and underlying-event effects.

We have showered the Les Houches events that we have produced using the
\RES{} code, employing the \HW angular-ordered shower (that we label as
``QTilde'' shower), and the default \PY{} shower with fully local
recoil~\cite{Cabouat:2017rzi} (that we denote as ``Dipole'' shower), as well
as the {\tt Pythia} implementation of the Vincia shower~\cite{Brooks:2020upa,
  Skands:2020lkd}.

For all the different types of shower we use the default setting of their
parameters, and we only change the charm mass to agree with the value we have
used to generate the $Wc$ sample (see Eq.(\ref{eq:phys_values})).
In Sect.~\ref{sec:detailsNLOPS} we provide additional details of the showers
we have employed, while in Sect.~\ref{sec:anatomy} we investigate the impact
of several ingredients provided by the general-purpose MC generators on the
simulation of $Wc$ events.  Finally, in Sect.~\ref{sec:CMS} we compare our
results against the experimental data at $\sqrt{s}=13$~GeV, taken by the CMS
Collaboration.

\subsection{The \RES matching with \HW and \PY}
\label{sec:detailsNLOPS}
\subsubsection{QTilde shower}
The relevant parameters to shower the POWHEG-generated events with the QTilde
shower are taken from the {\tt LHE-POWHEG.in} input card distributed in the
public release of the \HW{} code.  In particular the options
\begin{verbatim}
set /Herwig/Shower/ShowerHandler:MaxPtIsMuF Yes
set /Herwig/Shower/ShowerHandler:RestrictPhasespace Yes
\end{verbatim}
instruct the shower to veto all the emissions with transverse momentum larger
than the {\tt scalup} variable, which is the hardness of the POWHEG
emission.  We set the charm mass to $1.5$~GeV
\begin{verbatim}
set /Herwig/Particle/c:NominalMass 1.5*GeV
set /Herwig/Particle/cbar:NominalMass 1.5*GeV
\end{verbatim}
and we switch off the decay of unstable hadrons, in order to analyse more
quickly the generated events
\begin{verbatim}
set LesHouchesHandler:DecayHandler NULL
\end{verbatim}
By default, the QTilde shower includes also QED emissions, that can be
switched off with the option
\begin{verbatim}
set /Herwig/Shower/ShowerHandler:Interactions QCD
\end{verbatim}
The running coupling is evaluated at two loops in the
Catani-Marchesini-Webber scheme~\cite{Catani:1990rr}, with
$\as^{\rm \sss CMW}(\mZ)=0.1186$.

\subsubsection{Dipole shower}
We also considered the \PY{} implementation of a dipole shower with a fully
local recoil~\cite{Cabouat:2017rzi}
\begin{verbatim}
SpaceShower:dipoleRecoil = on
\end{verbatim}
We employed the {\tt PowhegHooks} class to veto emissions harder than the
POWHEG one, all the options correspond to the default ones, present in the
{\tt main31.cmnd} input card in the public release of the \PY{} code.  To
generate only the QCD shower, we set
\begin{verbatim}
SpaceShower:QEDshowerByQ = off      
SpaceShower:QEDshowerByL = off      
TimeShower:QEDshowerByQ = off        
TimeShower:QEDshowerByL = off
TimeShower:QEDshowerByGamma = off
TimeShower:QEDshowerByOther  = off
\end{verbatim}
and the hadrons decay is turned off with the flag
\begin{verbatim}
HadronLevel:Decay = off
\end{verbatim}
The running coupling is evaluated at one loop in the $\overline{\rm MS}$
scheme, and $\as^{\rm \sss \overline{MS}}(\mZ)=0.140$.

\subsubsection{Vincia shower}
At difference with respect to the Dipole shower, Vincia is an antenna
shower. To use its implementation in \PY, we set
\begin{verbatim}
PartonShowers:model = 2
\end{verbatim}
To properly include charm-mass effects, one has to include the option
\begin{verbatim}
Vincia:nFlavZeroMass = 3
\end{verbatim}
as, by default, this number is set to 4, meaning that only top and bottom are
treated as massive.  The QED shower can be turned off with the option
\begin{verbatim}
Vincia:ewMode = 0
\end{verbatim}
The running coupling is evaluated at two loops in the
Catani-Marchesini-Webber scheme, and $\as^{\rm\sss CMW}(\mZ)=0.127$.
Also in this case, we use the {\tt PowhegHooks} class to avoid to generate
emissions harder than the POWHEG one and we do not include the unstable
hadrons decay in our simulation.

\subsection{NLO + PS results with hadronization and  underlying-event effects}
\label{sec:anatomy}

In this section we discuss the impact of the parton shower, of
the hadronization and of the underlying events on the
results generated using our simulation in the \RES.  The results
presented in this section do not include the QED shower.

We focus on the $\mu^+\nu_{\mu}\bar{c}$ signature at $\sqrt{s}=13$~TeV.  We
use the values of the parameters described in Sect.~\ref{sec:params}, and the
B$_2$ setting of Table~\ref{tab:hdamp}, i.e.~the central values for the
renormalization and factorization scales are computed using the
underlying-Born kinematics, according to Eq.~(\ref{eq:H_T}), and as damping
function we use the $f_2$ profile function of Eq.~(\ref{eq:f2}) with $h=0.1$.

We begin by analyzing the effect of the parton shower on the \PWG
distributions~(PWG).  If, at the end of the shower, an event contains more
than one muon (or anti-muon) passing the charged leptonic cuts, only the
hardest one is considered to reconstruct the $W$ kinematics.

Since the aim of our analysis is to enhance the sensitivity of the measured
cross section to the strange quark PDF, we proceed as follows: if more
than one $c$ quark or charmed hadron is present, they are all considered,
with the exception of hadrons containing a $c\bar{c}$ pair, which are counted
as not charmed.
In addition, when in an event the charmed-quark and the muon charges have the
same sign~(SS), the weight of the event is subtracted from the differential
cross sections, while, when they have opposite sign~(OS), the weight is
added.
This procedure effectively removes the background due to events where the
final-state charm comes from gluon splitting $g\to c\bar{c}$, and not from an
initial-state strange quark emitting a $W$ boson.  Charmed quarks originated
from the gluon splitting have the same probability to be SS or OS, so their
contribution cancels, when the above procedure is applied.

Jets are reconstructed using the anti-$\kt$ algorithm with $R=0.5$. The charm
content of a jet is defined as $N_c-N_{\bar{c}}$, being $N_c$($N_{\bar{c}}$)
the number of $c$($\bar{c}$) quarks clustered in the jet, and the event
weight that we use to fill a jet-related distribution is further weighted by
the factor $\pm (N_c-N_{\bar{c}})$, where we use the $+$~sign when the muon
has negative charge, and the $-$~sign otherwise.

In the following, we consider both inclusive results (with the technical cut
on the $W$ boson virtuality of Eq.~(\ref{eq:tech_cut}) always in place), as
well as more exclusive ones, where we apply the following acceptance cuts
\begin{equation}
\label{eq:cutspheno}  
|\eta_{\mu}| < 2.4\,, \quad p_{\sss \perp \mu}>26~{\rm GeV}, \quad
\mWT>20~{\rm GeV}, \quad |\eta_c|<2.4\,, \quad p_{\sss \perp c}>5~{\rm GeV},
\end{equation}
used by the CMS Collaboration in their analyses.
\begin{figure}[htb]
\includegraphics[width=0.515\textwidth,page=8]{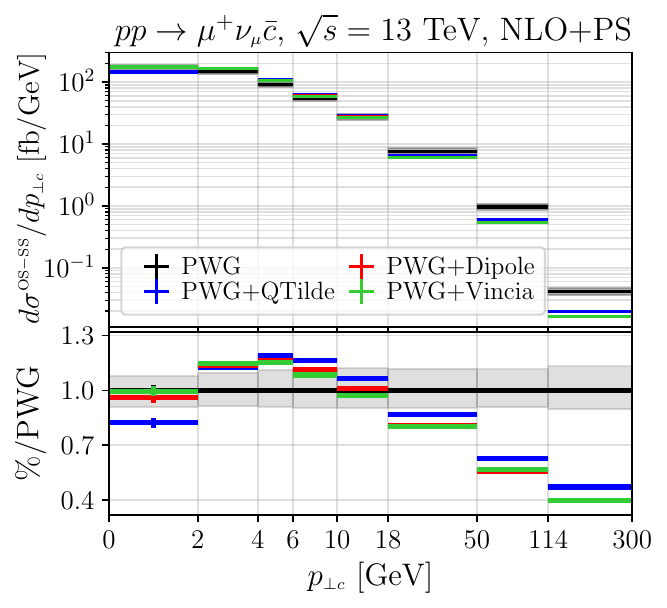}
\includegraphics[width=0.515\textwidth,page=6]{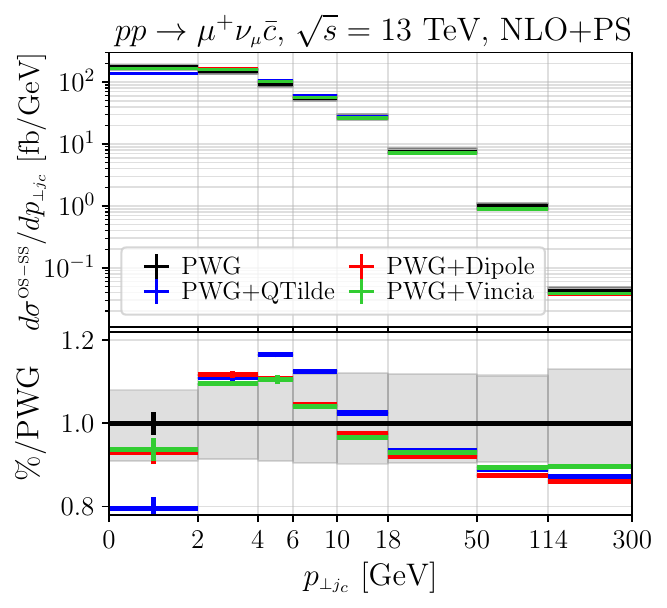}
\caption{  
  Effect of three different parton showers (QTilde in blue, Dipole in
  red and Vincia in green) on the events produced with the \RES~(PWG), in
  black, for the transverse momentum of the charmed quark~(left panel) and of
  the charmed jet~(right panel), for inclusive cuts. The grey band
  corresponds to the scale variations of the PWG distribution.
\label{fig:pTcinclPS}}
\end{figure}
At the inclusive level, the parton shower acts non-trivially on the
transverse-momentum distribution of the charm quark and of the charmed jet,
as portrayed in Fig.~\ref{fig:pTcinclPS}.  In particular, one can notice that
the hard tail is depleted, and the region around 5~GeV is instead increased.
The effect is milder for the $c$-jet, as radiation inside the cone with jet
radius $R=0.5$ is still collected in the jet, but more evident for the
hadron: in fact, in the high transverse-momentum region, the ratio of the LHE
results with the showered ones reaches values around 50\% for the
quark/hadron, and 10\% for the charmed jet.

\begin{figure}[htb]
\includegraphics[width=0.515\textwidth,page=8]{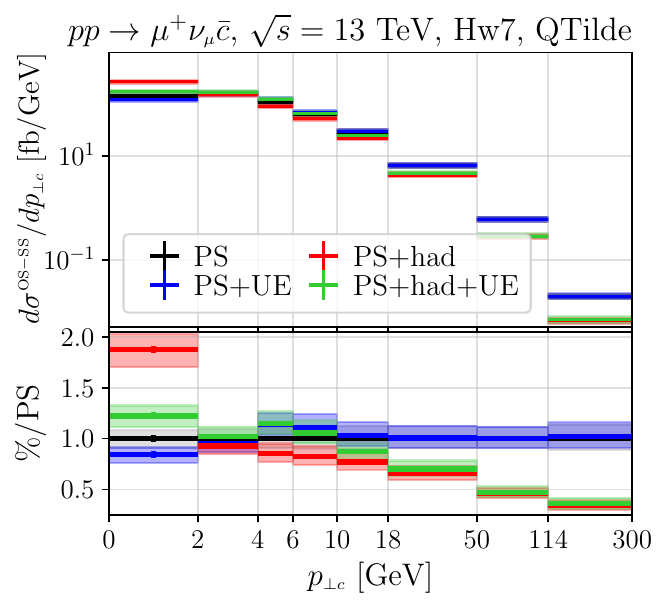}\hspace{-2.mm}
\includegraphics[width=0.515\textwidth,page=6]{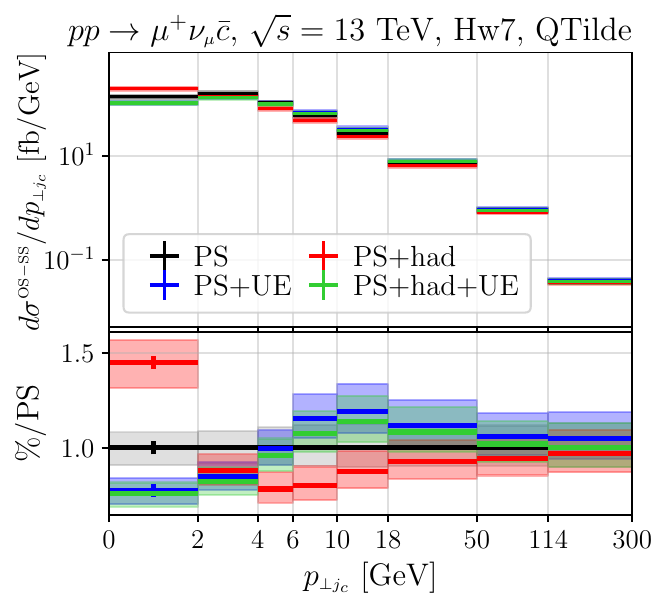}
\caption{ Effect of the underlying event~(UE) and of the hadronization~(had)
  on the NLO+PS distributions obtained with the \HW{} QTilde shower for the
  transverse-momentum distribution of the charmed quark~(left panel) and of
  the charmed jet~(right panel) for inclusive cuts. The grey band corresponds
  to renormalization- and factorization-scale variation, obtained using the
  standard reweighting procedure implemented in the \RES{} framework.
\label{fig:pTcinclHerwig}}
\end{figure}

\begin{figure}[htb]
\includegraphics[width=0.515\textwidth,page=8]{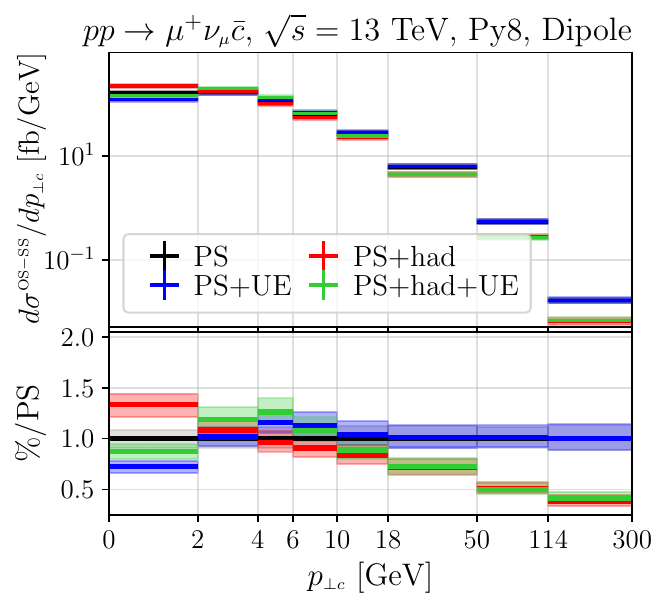}\hspace{-2.mm}
\includegraphics[width=0.515\textwidth,page=6]{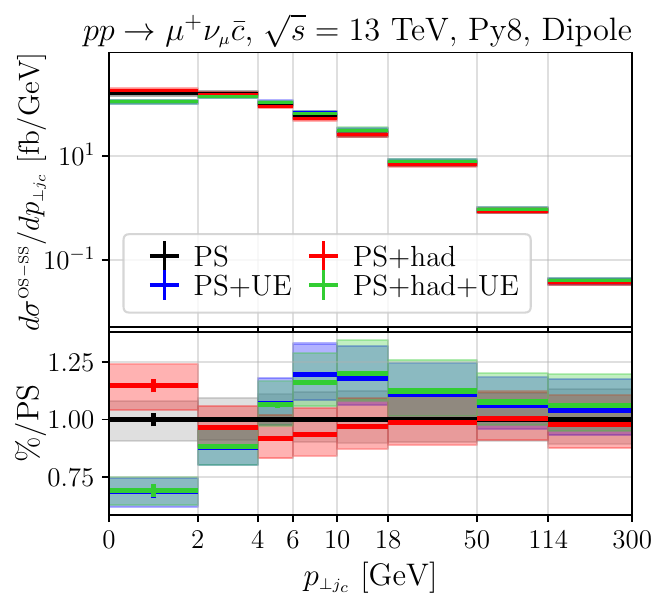}
\caption{Same as Fig.~\ref{fig:pTcinclHerwig}, but for the \PY{} Dipole shower.
\label{fig:pTcinclPythia}}
\end{figure}
In order to be able to compare theoretical predictions with data, we
also have to include hadronization corrections and the simulation of
the underlying-event~(UE) production.  In Figs.~\ref{fig:pTcinclHerwig}
and~\ref{fig:pTcinclPythia} we illustrate their effect on the NLO+PS distributions
obtained with the \HW{} QTilde shower, in
Fig.~\ref{fig:pTcinclHerwig}, and with the \PY{} Dipole shower, in
Fig.~\ref{fig:pTcinclPythia}. In all cases we notice that the UE have
a small impact on the charmed-quark distribution, except in the region
$p_{\sss \perp c} <10$~GeV.  On the other hand, the underlying event
hardens the $c$-jet distribution as it provides further particles that
can be clustered in the jet.  Hadronization corrections are large (up
to 50\%) and they soften the $p_{\sss \perp}$ of the charmed
quark/hadron. Similar, but milder effects are seen for the charmed
jet.

\begin{figure}[htb]
\includegraphics[width=0.515\textwidth,page=8]{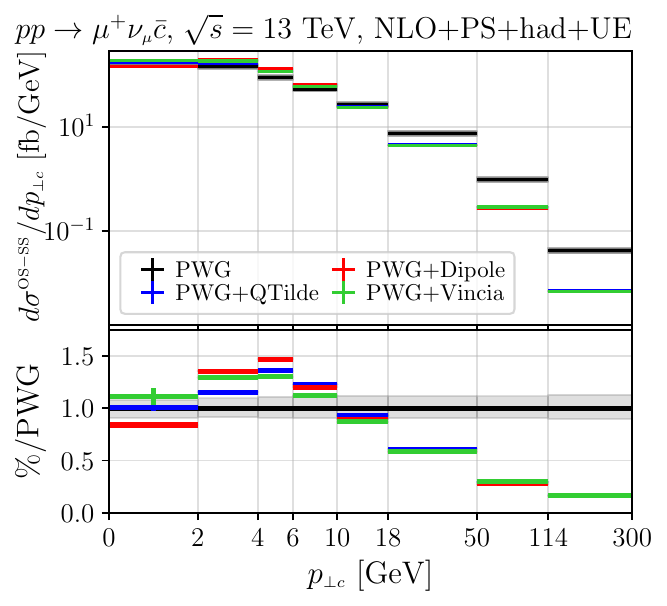}\hspace{-2.mm}
\includegraphics[width=0.515\textwidth,page=6]{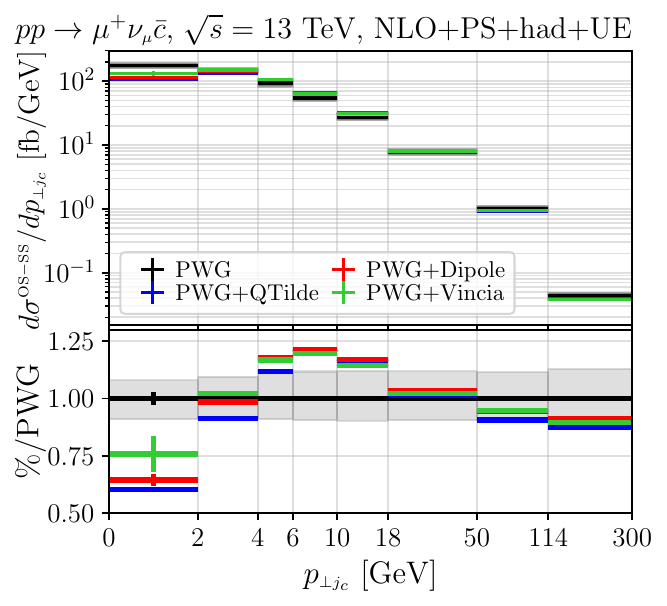}
\caption{Same as Fig.~\ref{fig:pTcinclPS}, but NLO+PS simulations have been
  supplemented with the underlying-event activity and the hadronization.
\label{fig:pTcinclfull}}
\end{figure}
In Fig.~\ref{fig:pTcinclfull} we then compare the full simulations
obtained with \HW{} and \PY{} against the PWG standalone results.
With the exception of the very low transverse-momentum region, \HW{}
and \PY{} distributions are in very good agreement.  In particular we
notice that now the charmed-hadron distributions, obtained with
all the three showers, are essentially indistinguishable for
$p_{\sss\perp c}>10$~GeV, while in Fig.~\ref{fig:pTcinclPS} the \HW{}
result was quite different from the \PY{} ones. The improved agreement
at the particle level is not unexpected, as the results without the
inclusion of the hadronization are very sensitive to the shower
cutoff, which is tuned alongside the hadronization
parameters.

\begin{figure}[htb]
\includegraphics[width=0.515\textwidth,page=11]{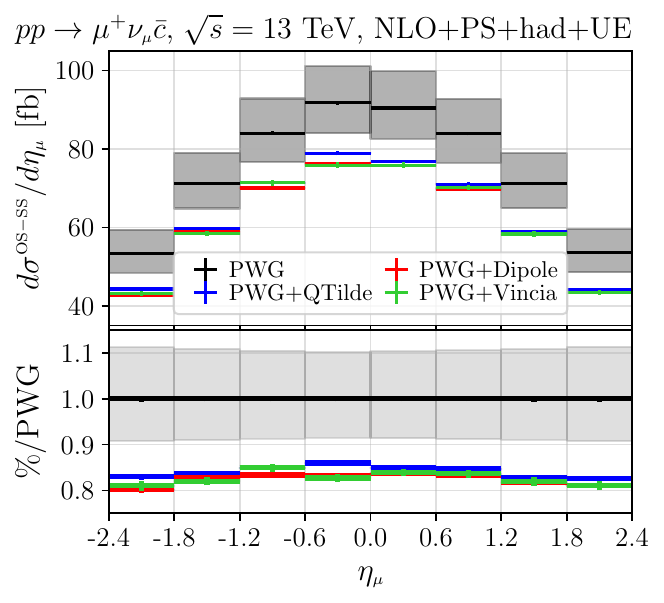}\hspace{-2.mm}
\includegraphics[width=0.515\textwidth,page=5]{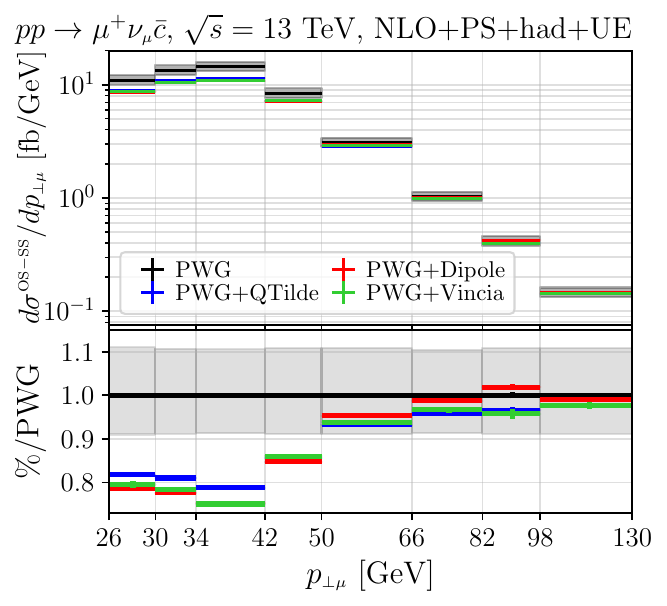}
\caption{Same as Fig.~\ref{fig:pTcinclfull} for the muon
  pseudo-rapidity~(left) and transverse momentum~(right), and using the cuts in Eq.~\eqref{eq:cutspheno}.
\label{fig:lepinclfull}}
\end{figure}
In Fig.~\ref{fig:lepinclfull} we plot the pseudo-rapidity\footnote{We remind
the reader that since we include lepton-mass effects when generating events,
as described in Sect.~\ref{sec:params}, the rapidity and pseudo-rapidity of
the leptons differ.} (left panel) and the transverse momentum (right panel)
of the muon, after applying the cuts in Eq.~\eqref{eq:cutspheno}.  Leptonic
distributions are indirectly affected by the shower, that induces a
modification of the momenta of the charmed quarks/hadrons.  In particular,
the softening of the charm quarks induced by the hadronization reduces the
cross section. This reduction is clearly visible on the left panel of
Fig.~\ref{fig:lepinclfull}, with an almost uniform reduction of the cross
section in the whole pseudo-rapidity range, of roughly 15\%, with respect to
the LHE result. In addition, all the three shower results are in very good
agreement with each others.
As far as the transverse momentum of the muon is concerned, shown on the
right panel, we observe a significant shape variation as the bulk of the
distribution, peaked around $p_{\sss \perp \mu} \approx \mW/2$, gets
depleted. Differences among the three showers amount to few percents,
and are much smaller than the scale-variation band.

\subsection{Comparison with the CMS data}
\label{sec:CMS}
In this section we compare our results against the CMS measurements at
$\sqrt{s}=13$~TeV for the associated production of a $W$ boson and a charm
quark~\cite{CMS:2018dxg}.  The $W$ boson is identified from its decay
products: only events with a muon with $p_{\sss \perp \mu} > 26$~GeV and
$|\eta_\mu|<2.4$ are selected.  The charm quarks are tagged by reconstructing
the $D^{*}(2010)$ mesons, which are required to have $p_{\sss\perp
  D^*}>5$~GeV and $|\eta_{\sss D^*}|<2.4$.\footnote{These cuts correspond to
the ones we have used in Sect.~\ref{sec:anatomy}, with the exception that
here no cut on $\mWT$ is imposed.}  The background arising from $g\to
c\bar{c}$ splitting is removed subtracting contributions where the charge of
the resolved $D^*$ meson and the muon have the same sign, as described in
Sect.~\ref{sec:anatomy}.  We employ the {\tt Rivet3}~\cite{Bierlich:2019rhm}
plugin to analyze the events generated by our simulations.  Since the $W$
boson is reconstructed from dressed leptons, in this case we also include the
effect of the QED shower.  All the other parameters are identical to those
presented in Sect.~\ref{sec:anatomy}.

\begin{figure}
\includegraphics[width=0.3334\textwidth, page=1]{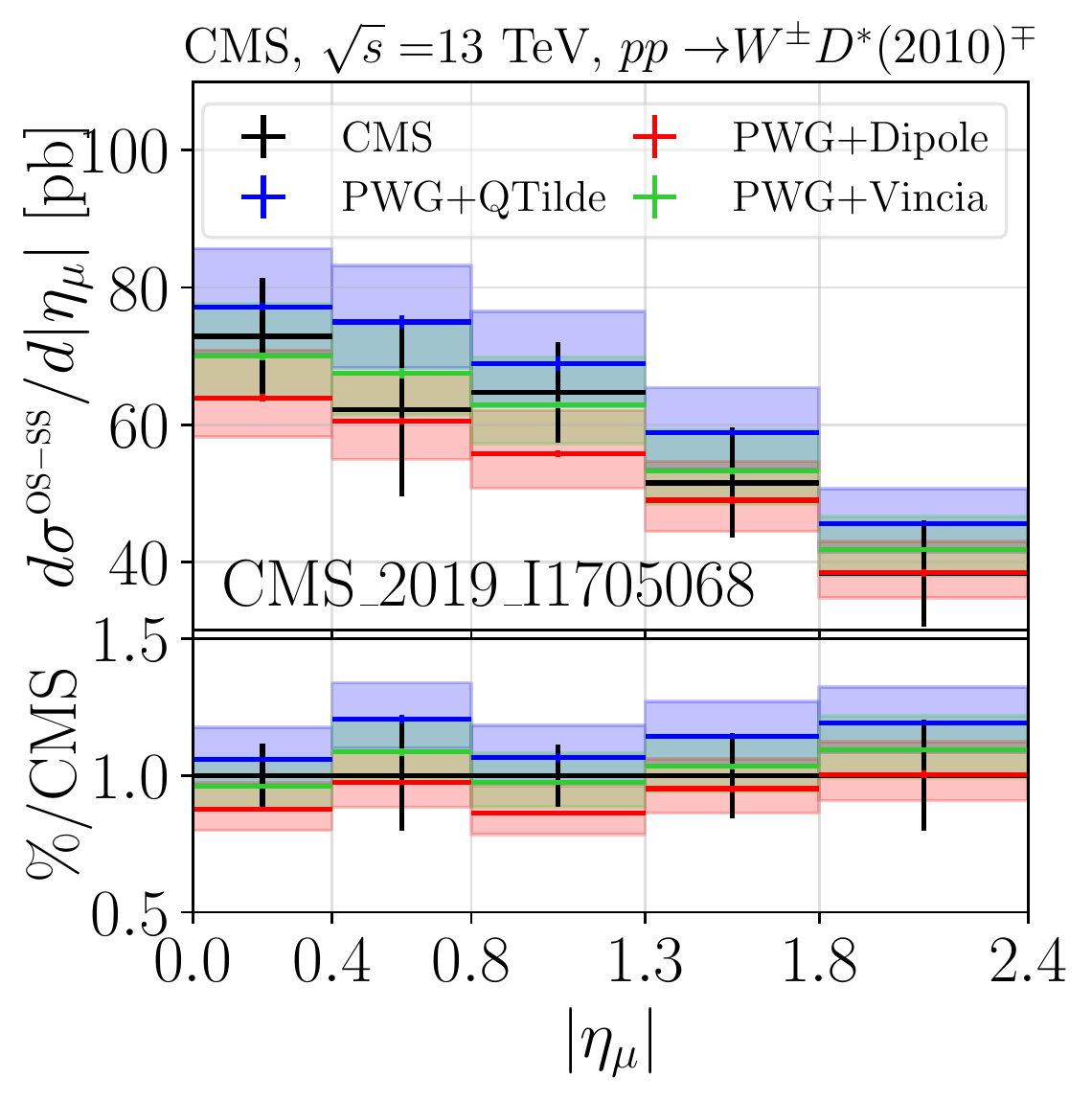}%
\includegraphics[width=0.3334\textwidth, page=2]{figs/CMS13TeV}%
\includegraphics[width=0.3334\textwidth, page=3]{figs/CMS13TeV}
\caption{ Muon pseudo-rapidity in $pp\to WD^*$ events at $\sqrt{s}=13$~TeV,
  using the selection cuts of the CMS analysis of Ref.~\cite{CMS:2018dxg},
  which are implemented in the Rivet analysis CMS\_2019\_I705068.  The
  experimental data and their combined statistical and systematic errors are
  drawn in black. The particle-level results obtained interfacing the \RES{}
  with the \HW{} (QTilde, in blue) and the \PY{} showers (Dipole, in red, and
  Vincia, in green) are also shown together with their scale-uncertainty
  bands, produced by varying $\muf$ and $\mur$ in the PWG calculation,
  performing the 7-point scale variations of Eq.~(\ref{eq:7-point}). 
 \label{fig:CMS}}
\end{figure}

The results of our comparison are illustrated in Fig.~\ref{fig:CMS} for the
inclusive $W^{\mp}D^*(2010)^{\pm}$ process, on the left panel, for $W^+
D^*(2010)^{-}$, on the middle panel, and for the $W^- D^*(2010)^{+}$
signature, on the right panel.  Among all the showers, Vincia seems to be
following the data more closely, slightly underestimating them, in the
default \PY{} configuration, while the \HW{} results do instead lead to a
larger value of the cross section.  We notice that the Dipole shower
distributions showed here are very similar to the ones presented in
Ref.~\cite{Bevilacqua:2021ovq}, where different PDF sets and tunes were
employed.  We want however to stress that we use the default tune of the
showers, and this distribution is particularly sensitive to the hadronization
modelling.  For example, the \HW{} hadronization parameters have been tuned
in Ref.~\cite{Bewick:2019rbu} using LEP data at the $Z$ mass. Due to the fact
that $B$ mesons primarily decay into $D$ ones, the charm hadronization is
further affected by its interplay with the bottom hadronization.  Thus, a
dedicated tune will certainly allow for a better comparison with the data.

\section{Conclusions}
In this article we have presented the implementation of a NLO + parton-shower
generator for massive-charm production in association with a
leptonically-decaying $W$ boson in hadron collisions in the \RES{} framework.
This process is particularly crucial for the determination of the
strange-quark content of the proton.

All spin-correlation effects and off-shell contributions in the $W$ decay are
properly included, and also non-diagonal elements in the
Cabibbo-Kobayashi-Maskawa matrix.

The leading-order and the one-loop matrix elements have been computed
analytically, allowing for a very fast and numerically-stable evaluation of
the amplitudes.

Using the flexibility of the POWHEG method on the possibility to separate the
real-radiation contribution into a finite and a divergent part, that is then
exponentiated in the Sudakov form factor, we have implemented different ways
to separate the real term and assessed the sensitivity of several kinematic
distributions to this extra degree of freedom.

The main source of theoretical uncertainty remains the dependence on the
factorization and renormalization scales, of the order of 10\% for most
kinematic distributions.

Events generated with our code have been interfaced with the \HW{} QTilde
shower, as well with the \PY{} default shower and the Vincia one.  We have
investigated the effect of the parton shower, hadronization and of the
underlying-event activity on several kinematic distributions.  In particular
hadronization corrections turned out to have the highest impact on the
differential cross sections.

Particle-level results obtained with the three shower models we have used
agree very well with each other, once hadronization corrections and the
underlying-event simulation are also included.

Finally we compared our results with the CMS data for the $WD^*$ signature,
finding good agreement, within the scale-variation bands.

\section*{Acknowledgments}
The fixed order NLO part of the code we have presented in this paper was
computed and implemented during S.F.R's Master thesis project.  S.F.R. wants
to acknowledge her co-advisors Federico Granata and Paolo Nason for guidance
during the implementation of the project.  S.F.R.~also wants to thank Rob
Verheyen for fruitful discussions on the Vincia shower.  S.F.R.~and
C.O.~thank Katerina Lipka for useful clarifications concerning the CMS Rivet
analysis, and Tom\'a\v{s} Je\v{z}o and Alexander Huss for useful comments on
the manuscript.


\providecommand{\href}[2]{#2}\begingroup\raggedright\endgroup


\begin{thebibliography}{10}

\bibitem{Baur:1993zd}
U.~Baur, F.~Halzen, S.~Keller, M.~L. Mangano and K.~Riesselmann, \emph{{The
  Charm content of $W$ + 1 jet events as a probe of the strange quark
  distribution function}},
  \href{http://dx.doi.org/10.1016/0370-2693(93)91553-Y}{\emph{Phys. Lett. B}
  {\bf 318} (1993) 544--548}, [\href{https://arxiv.org/abs/hep-ph/9308370}{{\tt
  hep-ph/9308370}}].

\bibitem{ATLAS:2014jkm}
{\scshape ATLAS} collaboration, G.~Aad et~al., \emph{{Measurement of the
  production of a $W$ boson in association with a charm quark in $pp$
  collisions at $\sqrt{s} =$ 7 TeV with the ATLAS detector}},
  \href{http://dx.doi.org/10.1007/JHEP05(2014)068}{\emph{JHEP} {\bf 05} (2014)
  068}, [\href{https://arxiv.org/abs/1402.6263}{{\tt 1402.6263}}].

\bibitem{ATLAS:2023ibp}
{\scshape ATLAS} collaboration, \emph{{Measurement of the production of a $W$
  boson in association with a charmed hadron in $pp$ collisions at $\sqrt{s} =
  13\,\mathrm{TeV}$ with the ATLAS detector}},
  \href{https://arxiv.org/abs/2302.00336}{{\tt 2302.00336}}.

\bibitem{CMS:2013wql}
{\scshape CMS} collaboration, S.~Chatrchyan et~al., \emph{{Measurement of
  Associated W + Charm Production in pp Collisions at $\sqrt{s}$ = 7 TeV}},
  \href{http://dx.doi.org/10.1007/JHEP02(2014)013}{\emph{JHEP} {\bf 02} (2014)
  013}, [\href{https://arxiv.org/abs/1310.1138}{{\tt 1310.1138}}].

\bibitem{CMS:2021oxn}
{\scshape CMS} collaboration, A.~Tumasyan et~al., \emph{{Measurements of the
  associated production of a W boson and a charm quark in
  proton\textendash{}proton collisions at $\sqrt{s}=8\,\text {TeV} $}},
  \href{http://dx.doi.org/10.1140/epjc/s10052-022-10897-7}{\emph{Eur. Phys. J.
  C} {\bf 82} (2022) 1094}, [\href{https://arxiv.org/abs/2112.00895}{{\tt
  2112.00895}}].

\bibitem{CMS:2018dxg}
{\scshape CMS} collaboration, A.~M. Sirunyan et~al., \emph{{Measurement of
  associated production of a W boson and a charm quark in proton-proton
  collisions at $\sqrt{s} =$ 13 TeV}},
  \href{http://dx.doi.org/10.1140/epjc/s10052-019-6752-1}{\emph{Eur. Phys. J.
  C} {\bf 79} (2019) 269}, [\href{https://arxiv.org/abs/1811.10021}{{\tt
  1811.10021}}].

\bibitem{LHCb:2015bwt}
{\scshape LHCb} collaboration, R.~Aaij et~al., \emph{{Study of $W$ boson
  production in association with beauty and charm}},
  \href{http://dx.doi.org/10.1103/PhysRevD.92.052001}{\emph{Phys. Rev. D} {\bf
  92} (2015) 052001}, [\href{https://arxiv.org/abs/1505.04051}{{\tt
  1505.04051}}].

\bibitem{Giele:1995kr}
W.~T. Giele, S.~Keller and E.~Laenen, \emph{{QCD corrections to $W$ boson plus
  heavy quark production at the Tevatron}},
  \href{http://dx.doi.org/10.1016/0370-2693(96)00078-0}{\emph{Phys. Lett. B}
  {\bf 372} (1996) 141--149}, [\href{https://arxiv.org/abs/hep-ph/9511449}{{\tt
  hep-ph/9511449}}].

\bibitem{Stirling:2012vh}
W.~J. Stirling and E.~Vryonidou, \emph{{Charm production in association with an
  electroweak gauge boson at the LHC}},
  \href{http://dx.doi.org/10.1103/PhysRevLett.109.082002}{\emph{Phys. Rev.
  Lett.} {\bf 109} (2012) 082002}, [\href{https://arxiv.org/abs/1203.6781}{{\tt
  1203.6781}}].

\bibitem{Czakon:2020coa}
M.~Czakon, A.~Mitov, M.~Pellen and R.~Poncelet, \emph{{NNLO QCD predictions for
  W+c-jet production at the LHC}},
  \href{http://dx.doi.org/10.1007/JHEP06(2021)100}{\emph{JHEP} {\bf 06} (2021)
  100}, [\href{https://arxiv.org/abs/2011.01011}{{\tt 2011.01011}}].

\bibitem{Czakon:2022khx}
M.~Czakon, A.~Mitov, M.~Pellen and R.~Poncelet, \emph{{A detailed investigation
  of W+c-jet at the LHC}},
  \href{http://dx.doi.org/10.1007/JHEP02(2023)241}{\emph{JHEP} {\bf 02} (2023)
  241}, [\href{https://arxiv.org/abs/2212.00467}{{\tt 2212.00467}}].

\bibitem{Bevilacqua:2021ovq}
G.~Bevilacqua, M.~V. Garzelli, A.~Kardos and L.~Toth, \emph{{W+charm production
  with massive c quarks in PowHel}},
  \href{https://arxiv.org/abs/2106.11261}{{\tt 2106.11261}}.

\bibitem{Nason:2004rx}
P.~Nason, \emph{{A New method for combining NLO QCD with shower Monte Carlo
  algorithms}},
  \href{http://dx.doi.org/10.1088/1126-6708/2004/11/040}{\emph{JHEP} {\bf 11}
  (2004) 040}, [\href{https://arxiv.org/abs/hep-ph/0409146}{{\tt
  hep-ph/0409146}}].

\bibitem{Frixione:2002ik}
S.~Frixione and B.~R. Webber, \emph{{Matching NLO QCD computations and parton
  shower simulations}},
  \href{http://dx.doi.org/10.1088/1126-6708/2002/06/029}{\emph{JHEP} {\bf 06}
  (2002) 029}, [\href{https://arxiv.org/abs/hep-ph/0204244}{{\tt
  hep-ph/0204244}}].

\bibitem{Alwall:2014hca}
J.~Alwall, R.~Frederix, S.~Frixione, V.~Hirschi, F.~Maltoni, O.~Mattelaer
  et~al., \emph{{The automated computation of tree-level and next-to-leading
  order differential cross sections, and their matching to parton shower
  simulations}}, \href{http://dx.doi.org/10.1007/JHEP07(2014)079}{\emph{JHEP}
  {\bf 07} (2014) 079}, [\href{https://arxiv.org/abs/1405.0301}{{\tt
  1405.0301}}].

\bibitem{Nason:2020lxx}
P.~Nason, C.~Oleari, M.~Rocco and M.~Zaro, \emph{{An interface between the
  POWHEG BOX and MadGraph5$\_$aMC@NLO}},
  \href{http://dx.doi.org/10.1140/epjc/s10052-020-08559-7}{\emph{Eur. Phys. J.
  C} {\bf 80} (2020) 985}, [\href{https://arxiv.org/abs/2008.06364}{{\tt
  2008.06364}}].

\bibitem{Frixione:2007vw}
S.~Frixione, P.~Nason and C.~Oleari, \emph{{Matching NLO QCD computations with
  Parton Shower simulations: the POWHEG method}},
  \href{http://dx.doi.org/10.1088/1126-6708/2007/11/070}{\emph{JHEP} {\bf 11}
  (2007) 070}, [\href{https://arxiv.org/abs/0709.2092}{{\tt 0709.2092}}].

\bibitem{Alioli:2010xd}
S.~Alioli, P.~Nason, C.~Oleari and E.~Re, \emph{{A general framework for
  implementing NLO calculations in shower Monte Carlo programs: the POWHEG
  BOX}}, \href{http://dx.doi.org/10.1007/JHEP06(2010)043}{\emph{JHEP} {\bf 06}
  (2010) 043}, [\href{https://arxiv.org/abs/1002.2581}{{\tt 1002.2581}}].

\bibitem{Jezo:2015aia}
T.~Je\v{z}o and P.~Nason, \emph{{On the Treatment of Resonances in
  Next-to-Leading Order Calculations Matched to a Parton Shower}},
  \href{http://dx.doi.org/10.1007/JHEP12(2015)065}{\emph{JHEP} {\bf 12} (2015)
  065}, [\href{https://arxiv.org/abs/1509.09071}{{\tt 1509.09071}}].

\bibitem{Barze:2012tt}
L.~Barze, G.~Montagna, P.~Nason, O.~Nicrosini and F.~Piccinini,
  \emph{{Implementation of electroweak corrections in the POWHEG BOX: single W
  production}}, \href{http://dx.doi.org/10.1007/JHEP04(2012)037}{\emph{JHEP}
  {\bf 04} (2012) 037}, [\href{https://arxiv.org/abs/1202.0465}{{\tt
  1202.0465}}].

\bibitem{Mazzitelli:2020jio}
J.~Mazzitelli, P.~F. Monni, P.~Nason, E.~Re, M.~Wiesemann and G.~Zanderighi,
  \emph{{Next-to-Next-to-Leading Order Event Generation for Top-Quark Pair
  Production}},
  \href{http://dx.doi.org/10.1103/PhysRevLett.127.062001}{\emph{Phys. Rev.
  Lett.} {\bf 127} (2021) 062001},
  [\href{https://arxiv.org/abs/2012.14267}{{\tt 2012.14267}}].

\bibitem{Mazzitelli:2021mmm}
J.~Mazzitelli, P.~F. Monni, P.~Nason, E.~Re, M.~Wiesemann and G.~Zanderighi,
  \emph{{Top-pair production at the LHC with MINNLO$_{PS}$}},
  \href{http://dx.doi.org/10.1007/JHEP04(2022)079}{\emph{JHEP} {\bf 04} (2022)
  079}, [\href{https://arxiv.org/abs/2112.12135}{{\tt 2112.12135}}].

\bibitem{Zanoli:2021iyp}
S.~Zanoli, M.~Chiesa, E.~Re, M.~Wiesemann and G.~Zanderighi,
  \emph{{Next-to-next-to-leading order event generation for VH production with
  H \textrightarrow{}$ b\overline{b} $ decay}},
  \href{http://dx.doi.org/10.1007/JHEP07(2022)008}{\emph{JHEP} {\bf 07} (2022)
  008}, [\href{https://arxiv.org/abs/2112.04168}{{\tt 2112.04168}}].

\bibitem{Gavardi:2022ixt}
A.~Gavardi, C.~Oleari and E.~Re, \emph{{NNLO+PS Monte Carlo simulation of
  photon pair production with MiNNLO$_{PS}$}},
  \href{http://dx.doi.org/10.1007/JHEP09(2022)061}{\emph{JHEP} {\bf 09} (2022)
  061}, [\href{https://arxiv.org/abs/2204.12602}{{\tt 2204.12602}}].

\bibitem{Monni:2019whf}
P.~F. Monni, P.~Nason, E.~Re, M.~Wiesemann and G.~Zanderighi,
  \emph{{MiNNLO$_{PS}$: a new method to match NNLO QCD to parton showers}},
  \href{http://dx.doi.org/10.1007/JHEP05(2020)143}{\emph{JHEP} {\bf 05} (2020)
  143}, [\href{https://arxiv.org/abs/1908.06987}{{\tt 1908.06987}}].

\bibitem{Monni:2020nks}
P.~F. Monni, E.~Re and M.~Wiesemann, \emph{{MiNNLO$_{\text {PS}}$: optimizing
  $2\rightarrow 1$ hadronic processes}},
  \href{http://dx.doi.org/10.1140/epjc/s10052-020-08658-5}{\emph{Eur. Phys. J.
  C} {\bf 80} (2020) 1075}, [\href{https://arxiv.org/abs/2006.04133}{{\tt
  2006.04133}}].

\bibitem{Cullen:2014yla}
G.~Cullen et~al., \emph{{G$\scriptsize{O}$S$\scriptsize{AM}$-2.0: a tool for
  automated one-loop calculations within the Standard Model and beyond}},
  \href{http://dx.doi.org/10.1140/epjc/s10052-014-3001-5}{\emph{Eur. Phys. J.
  C} {\bf 74} (2014) 3001}, [\href{https://arxiv.org/abs/1404.7096}{{\tt
  1404.7096}}].

\bibitem{Cullen:2011ac}
G.~Cullen, N.~Greiner, G.~Heinrich, G.~Luisoni, P.~Mastrolia, G.~Ossola et~al.,
  \emph{{Automated One-Loop Calculations with GoSam}},
  \href{http://dx.doi.org/10.1140/epjc/s10052-012-1889-1}{\emph{Eur. Phys. J.
  C} {\bf 72} (2012) 1889}, [\href{https://arxiv.org/abs/1111.2034}{{\tt
  1111.2034}}].

\bibitem{Alwall:2007st}
J.~Alwall, P.~Demin, S.~de~Visscher, R.~Frederix, M.~Herquet, F.~Maltoni
  et~al., \emph{{MadGraph/MadEvent v4: The New Web Generation}},
  \href{http://dx.doi.org/10.1088/1126-6708/2007/09/028}{\emph{JHEP} {\bf 09}
  (2007) 028}, [\href{https://arxiv.org/abs/0706.2334}{{\tt 0706.2334}}].

\bibitem{Campbell:2012am}
J.~M. Campbell, R.~K. Ellis, R.~Frederix, P.~Nason, C.~Oleari and C.~Williams,
  \emph{{NLO Higgs Boson Production Plus One and Two Jets Using the POWHEG BOX,
  MadGraph4 and MCFM}},
  \href{http://dx.doi.org/10.1007/JHEP07(2012)092}{\emph{JHEP} {\bf 07} (2012)
  092}, [\href{https://arxiv.org/abs/1202.5475}{{\tt 1202.5475}}].

\bibitem{Collins:1978wz}
J.~C. Collins, F.~Wilczek and A.~Zee, \emph{{Low-Energy Manifestations of Heavy
  Particles: Application to the Neutral Current}},
  \href{http://dx.doi.org/10.1103/PhysRevD.18.242}{\emph{Phys. Rev. D} {\bf 18}
  (1978) 242}.

\bibitem{Appelquist:1974tg}
T.~Appelquist and J.~Carazzone, \emph{{Infrared Singularities and Massive
  Fields}}, \href{http://dx.doi.org/10.1103/PhysRevD.11.2856}{\emph{Phys. Rev.
  D} {\bf 11} (1975) 2856}.

\bibitem{Cacciari:1998it}
M.~Cacciari, M.~Greco and P.~Nason, \emph{{The $p_{\rm T}$ spectrum in heavy
  flavor hadroproduction}},
  \href{http://dx.doi.org/10.1088/1126-6708/1998/05/007}{\emph{JHEP} {\bf 05}
  (1998) 007}, [\href{https://arxiv.org/abs/hep-ph/9803400}{{\tt
  hep-ph/9803400}}].

\bibitem{Luisoni:2015mpa}
G.~Luisoni, C.~Oleari and F.~Tramontano, \emph{{$ Wb\overline{b}j $ production
  at NLO with POWHEG+MiNLO}},
  \href{http://dx.doi.org/10.1007/JHEP04(2015)161}{\emph{JHEP} {\bf 04} (2015)
  161}, [\href{https://arxiv.org/abs/1502.01213}{{\tt 1502.01213}}].

\bibitem{Buckley:2014ana}
A.~Buckley, J.~Ferrando, S.~Lloyd, K.~Nordstr\"om, B.~Page, M.~R\"ufenacht
  et~al., \emph{{LHAPDF6: parton density access in the LHC precision era}},
  \href{http://dx.doi.org/10.1140/epjc/s10052-015-3318-8}{\emph{Eur. Phys. J.
  C} {\bf 75} (2015) 132}, [\href{https://arxiv.org/abs/1412.7420}{{\tt
  1412.7420}}].

\bibitem{NNPDF:2017mvq}
{\scshape NNPDF} collaboration, R.~D. Ball et~al., \emph{{Parton distributions
  from high-precision collider data}},
  \href{http://dx.doi.org/10.1140/epjc/s10052-017-5199-5}{\emph{Eur. Phys. J.
  C} {\bf 77} (2017) 663}, [\href{https://arxiv.org/abs/1706.00428}{{\tt
  1706.00428}}].

\bibitem{Campbell:2019dru}
J.~Campbell and T.~Neumann, \emph{{Precision Phenomenology with MCFM}},
  \href{http://dx.doi.org/10.1007/JHEP12(2019)034}{\emph{JHEP} {\bf 12} (2019)
  034}, [\href{https://arxiv.org/abs/1909.09117}{{\tt 1909.09117}}].

\bibitem{Campbell:2015qma}
J.~M. Campbell, R.~K. Ellis and W.~T. Giele, \emph{{A Multi-Threaded Version of
  MCFM}}, \href{http://dx.doi.org/10.1140/epjc/s10052-015-3461-2}{\emph{Eur.
  Phys. J. C} {\bf 75} (2015) 246},
  [\href{https://arxiv.org/abs/1503.06182}{{\tt 1503.06182}}].

\bibitem{Bellm:2016rhh}
J.~Bellm, G.~Nail, S.~Pl\"atzer, P.~Schichtel and A.~Si\'odmok, \emph{{Parton
  Shower Uncertainties with Herwig 7: Benchmarks at Leading Order}},
  \href{http://dx.doi.org/10.1140/epjc/s10052-016-4506-x}{\emph{Eur. Phys. J.
  C} {\bf 76} (2016) 665}, [\href{https://arxiv.org/abs/1605.01338}{{\tt
  1605.01338}}].

\bibitem{Cacciari:2011ma}
M.~Cacciari, G.~P. Salam and G.~Soyez, \emph{{FastJet User Manual}},
  \href{http://dx.doi.org/10.1140/epjc/s10052-012-1896-2}{\emph{Eur. Phys. J.
  C} {\bf 72} (2012) 1896}, [\href{https://arxiv.org/abs/1111.6097}{{\tt
  1111.6097}}].

\bibitem{Cacciari:2008gp}
M.~Cacciari, G.~P. Salam and G.~Soyez, \emph{{The anti-$k_t$ jet clustering
  algorithm}},
  \href{http://dx.doi.org/10.1088/1126-6708/2008/04/063}{\emph{JHEP} {\bf 04}
  (2008) 063}, [\href{https://arxiv.org/abs/0802.1189}{{\tt 0802.1189}}].

\bibitem{Bahr:2008pv}
M.~Bahr et~al., \emph{{Herwig++ Physics and Manual}},
  \href{http://dx.doi.org/10.1140/epjc/s10052-008-0798-9}{\emph{Eur. Phys. J.
  C} {\bf 58} (2008) 639--707}, [\href{https://arxiv.org/abs/0803.0883}{{\tt
  0803.0883}}].

\bibitem{Bellm:2019zci}
J.~Bellm et~al., \emph{{Herwig 7.2 release note}},
  \href{http://dx.doi.org/10.1140/epjc/s10052-020-8011-x}{\emph{Eur. Phys. J.
  C} {\bf 80} (2020) 452}, [\href{https://arxiv.org/abs/1912.06509}{{\tt
  1912.06509}}].

\bibitem{Sjostrand:2006za}
T.~Sjostrand, S.~Mrenna and P.~Z. Skands, \emph{{PYTHIA 6.4 Physics and
  Manual}}, \href{http://dx.doi.org/10.1088/1126-6708/2006/05/026}{\emph{JHEP}
  {\bf 05} (2006) 026}, [\href{https://arxiv.org/abs/hep-ph/0603175}{{\tt
  hep-ph/0603175}}].

\bibitem{Sjostrand:2014zea}
T.~Sj\"ostrand, S.~Ask, J.~R. Christiansen, R.~Corke, N.~Desai, P.~Ilten
  et~al., \emph{{An introduction to PYTHIA 8.2}},
  \href{http://dx.doi.org/10.1016/j.cpc.2015.01.024}{\emph{Comput. Phys.
  Commun.} {\bf 191} (2015) 159--177},
  [\href{https://arxiv.org/abs/1410.3012}{{\tt 1410.3012}}].

\bibitem{Cabouat:2017rzi}
B.~Cabouat and T.~Sj\"ostrand, \emph{{Some Dipole Shower Studies}},
  \href{http://dx.doi.org/10.1140/epjc/s10052-018-5645-z}{\emph{Eur. Phys. J.
  C} {\bf 78} (2018) 226}, [\href{https://arxiv.org/abs/1710.00391}{{\tt
  1710.00391}}].

\bibitem{Brooks:2020upa}
H.~Brooks, C.~T. Preuss and P.~Skands, \emph{{Sector Showers for Hadron
  Collisions}}, \href{http://dx.doi.org/10.1007/JHEP07(2020)032}{\emph{JHEP}
  {\bf 07} (2020) 032}, [\href{https://arxiv.org/abs/2003.00702}{{\tt
  2003.00702}}].

\bibitem{Skands:2020lkd}
P.~Skands and R.~Verheyen, \emph{{Multipole photon radiation in the Vincia
  parton shower}},
  \href{http://dx.doi.org/10.1016/j.physletb.2020.135878}{\emph{Phys. Lett. B}
  {\bf 811} (2020) 135878}, [\href{https://arxiv.org/abs/2002.04939}{{\tt
  2002.04939}}].

\bibitem{Catani:1990rr}
S.~Catani, B.~R. Webber and G.~Marchesini, \emph{{QCD coherent branching and
  semiinclusive processes at large x}},
  \href{http://dx.doi.org/10.1016/0550-3213(91)90390-J}{\emph{Nucl. Phys. B}
  {\bf 349} (1991) 635--654}.

\bibitem{Bierlich:2019rhm}
C.~Bierlich et~al., \emph{{Robust Independent Validation of Experiment and
  Theory: Rivet version 3}},
  \href{http://dx.doi.org/10.21468/SciPostPhys.8.2.026}{\emph{SciPost Phys.}
  {\bf 8} (2020) 026}, [\href{https://arxiv.org/abs/1912.05451}{{\tt
  1912.05451}}].

\bibitem{Bewick:2019rbu}
G.~Bewick, S.~Ferrario~Ravasio, P.~Richardson and M.~H. Seymour,
  \emph{{Logarithmic accuracy of angular-ordered parton showers}},
  \href{http://dx.doi.org/10.1007/JHEP04(2020)019}{\emph{JHEP} {\bf 04} (2020)
  019}, [\href{https://arxiv.org/abs/1904.11866}{{\tt 1904.11866}}].

\end{thebibliography}
\end{document}